\documentclass[useAMS,usenatbib]{mn2e}

\usepackage{graphicx}
\usepackage{psfig}
\usepackage{amssymb,amsmath}
\usepackage{mathrsfs}
\usepackage{times}
\usepackage{setspace}
\usepackage{color}

\def\aj{AJ}%
\def\araa{ARA\&A}%
\def\apj{ApJ}%
\def\apjl{ApJ}%
\def\apjs{ApJS}%
\def\apss{Ap\&SS}%
\def\aap{A\&A}%
\def\mnras{MNRAS}%
\def\pasp{PASP}%
\def\nat{Nature}%
\newcommand{\begit}{\begin{itemize}}
\newcommand{\enit}{\end{itemize}}
\newcommand{\begen}{\begin{enumerate}}
\newcommand{\enen}{\end{enumerate}}

\newcommand{\beq}{\begin{equation}}
\newcommand{\eeq}{\end{equation}}
\newcommand{\beqa}{\begin{eqnarray}} 
\newcommand{\eeqa}{\end{eqnarray}} 
\def\lesssim{\mathrel{\hbox{\rlap{\hbox{\lower5pt\hbox{$\sim$}}}\hbox{$<$}}}}
\def\gtrsim{\mathrel{\hbox{\rlap{\hbox{\lower5pt\hbox{$\sim$}}}\hbox{$>$}}}}

\title[Inhibiting Giant Planet Formation in Star Clusters]{Gas Giants in Hot Water: Inhibiting Giant Planet
Formation \& Planet Habitability in Dense Star Clusters Through Cosmic Time}

\author[Thompson]{Todd A.~Thompson \\
Department of Astronomy and Center for Cosmology \& Astro-Particle Physics
The Ohio State University, Columbus, Ohio 43210, USA\\
email: thompson@astronomy.ohio-state.edu}

\begin{document}
\voffset -1.5cm
\maketitle
\label{firstpage}
\begin{abstract}
I show that the temperature of nuclear star 
clusters, starburst clusters in M82, compact high-$z$ galaxies,
and some globular clusters of the Galaxy likely exceeded the ice line temperature 
($T_{\rm Ice}\approx150-170$\,K) during formation for a time comparable to the planet formation
timescale.  The protoplanetary disks within these systems 
will thus not have an ice line, decreasing the total material available for building 
protoplanetary embryos, inhibiting the 
formation of gas- and ice-giants if they form by core
accretion, and prohibiting habitability. Planet formation by gravitational instability is 
similarly suppressed because Toomre's $Q>1$ in all but
the most massive disks. I discuss these results in the context of 
the observed lack of planets in 47 Tuc. I predict that a similar 
search for planets in the globular cluster NGC 6366 (${\rm [Fe/H]}=-0.82$) should 
yield detections, whereas (counterintuitively) 
the relatively {\it metal-rich} globular clusters  NGC 6440, 6441, and 6388
should be devoid of giant planets.  The characteristic stellar surface density above
which $T_{\rm Ice}$ is exceeded in star clusters is $\sim6\times10^3\,\,{\rm M_\odot\,\,pc^{-2}}\,\,f_{\rm dg,\,MW}^{-1/2}$,
where $f_{\rm dg,\,MW}$ is the dust-to-gas ratio of the embedding material, normalized to the 
Milky Way value.  Simple estimates suggest that $\sim5-50$\% of the stars in the universe
formed in an environment exceeding this surface density.
Caveats and uncertainties are detailed.

\end{abstract}


\section{Introduction}
\label{section:introduction}

A key question is whether planet formation over cosmic time is concomitant with star formation,
and whether or not the efficiency of planet formation is independent of 
environment and formation epoch.  The correlation observed between giant planet 
frequency and host star metallicity by Gonzalez (1997)\footnote{See also Laughlin 2000, 
Gonzalez et al.~2001,
Laws et al.~2003, Santos et al.~2001, 2004, and Fischer \& Valenti 2005.} indicates 
that it is not, and implies that the $z\sim0$ planetary mass density $\Omega_p$ is at 
minimum a convolution of the enrichment history of the ISM of galaxies and 
their star formation rates.

However, the relationship between $\Omega_p$ and the $z=0$ stellar mass density 
$\Omega_\star$ is likely more complicated than a simple function of metallicity.
Much of the star formation in normal spiral galaxies, starbursts, 
and the  rapidly star-forming galaxies at high redshift 
occurs in massive star clusters (Lada \& Lada 2003; Portegies-Zwart et al.~2010; 
Kruijssen 2012). The dense environment accompanying stellar birth in these systems is 
likely to significantly impact the physics of planet formation.
For example, dynamical interactions, UV photoevaporation, and 
supernovae have all been suggested as mechanisms that might 
inhibit, disrupt, or affect the formation, migration, and survival of planets in 
dense stellar environments (Laughlin \& Adams 1998;
Armitage 2000; Adams et al.~2004; Fregeau et al.~2006;
Adams et al.~2006; Fatuzzo \& Adams 2008; 
Gorti \& Hollenbach 2009; Gorti et al.~2009; 
Proszkow \& Adams 2009; Spurzem et al.~2009; Olczak et al.~2012).  

Because the timescale for these processes to affect planet formation 
is comparable to the planet formation timescale itself 
($\sim$\,Myr), and because most stars form in clusters,
the planet formation 
efficiency is likely a function of the stellar density of the birth environment.
Since the bulk of the universe's stars are formed at $z\sim1-2$,
where the conditions for producing dense star clusters with 
high surface densities were common,
but metallicities were generally lower, the functional dependence of
$\Omega_p/\Omega_\star(z)$ may be complicated.

These issues are of relevance now because observations over the last 
decade have begun to constrain the incidence of planets in the local 
field disk population, in  
both open and globular clusters of the Galaxy, and, through microlensing,
the incidence of planets in the Milky Way's bulge.  These various data sets
and future extensions have provided and will provide a wealth of information
on the (likely) multi-dimensional space of planet formation efficiency as a 
function of birth environment.  In particular, 
Gilliland et al.~(2000) and Weldrake et al.~(2005) find a statistically 
significant lack of  ``hot Jupiters'' in the globular cluster 47 Tuc.  
A multitude of transit surveys in open clusters have begun to constrain
the efficiency of hot Jupiter formation in these less dense systems
(Mochejska et al.~2002, 2004, 2005, 2006; Hartman et al.~2005; Hidas et al.~2005;
Rosvick \& Robb 2006; Burke et al.~2006; Montalto et al.~2007; Hartman et al.~2009;
Nascimbeni et al.~2012; see the analysis in Van Saders \& Gaudi 2011),
and RV surveys have provided tighter constraints on the population
and, recently, some detections (Cochran et al.~2002; Paulson et al.~2004;
Quinn et al.~2012; Pasquini et al.~2012).  As I emphasize here and throughout
this paper, such surveys, if carried out for a number of globular and open  clusters
with a variety of metallicities and stellar densities, might provide the 
first quantitative measure of the planet formation efficiency as a function of 
both metallicity and stellar surface density.  

Complimentary to transit and RV surveys is microlensing, which could in
principle measure the planet frequency in the Galactic bulge via self-lensing
(Kriaga \& Paczynski 1994) relative to the frequency in the foreground disk
(e.g., Han \& Gould 1995).  Microlensing may also be used to constrain the 
planet frequency via self-lensing in M31 (Baltz \& Gondolo 2001; Chung et al.~2006).
Since the Galactic bulge has had a different formation and enrichment 
history than the disk or halo, one would expect a different planet formation 
efficiency per unit star formed.  Studies of the low mass IMF in the Galaxy's 
bulge and disk reveal a large difference in slope (see Fig.~8 from Zoccali et al.~2000)
that might manifest an underlying difference in the relative planet population.
Future observations of the relative planet frequency in these two environments
(bulge and disk) might be interpretable with constraints on the planet population
in the many different open and globular clusters of the Galaxy,
again with a range of metallicities and surface densities.
These studies would be further complemented by searches for the NIR signatures
of protoplanetary disks in young and dense stellar environments, such as 
in Arches and NGC 3603 (Stolte et al.~2004, 2010).

With this backdrop, here I consider the importance of the strong infrared 
irradiation expected in a newly-born star cluster.  Different from works 
focusing on photo-evaporation of protoplanetary disks, I show that the UV and optical emission from 
massive stars, which is efficiently absorbed and scattered by dust grains, creates a 
thermal bath similar to a hohlraum.   This strong thermal cluster irradiation
is the focus of this work.

I show that the heating and 
thermal structure of protoplanetary 
disks around Solar-type stars during the first few Myr of cluster evolution 
is in fact dominated by the irradiation from the cluster as a whole,
and not from the host star, for semi-major axes larger than a few AU.  This
keeps the disks essentially isothermal and hot as a function of radius, affecting 
their structure and their potential to form planets via gravitational instability
or core accretion.  In fact, I show that 
in the interior of massive embedded star clusters, the  temperature 
approaches and exceeds the condensation temperature of water ice, $T_{\rm Ice}\simeq150-170$\,K
(Podolak \& Zucker 2004; Lecar et al.~2006).  
Because the disks equilibrate to midplane temperatures at or above the value
corresponding to the incident flux, 
this means that protoplanetary disks in such environments have no ice 
line.  Stated succinctly, the central temperatures of many star clusters 
in formation exceed the fundamental temperature scale for planet formation 
by core-accretion: $T_{\rm Ice}$.  This
decreases the total amount of condensable material by a factor of $\sim3-5$, and thus
inhibits giant planet formation (Lodders 2003; Lecar et al.~2006).  Over a wider and less extreme range of 
cluster properties, the temperature is still very high, and this should in general
suppress planet formation by gravitational instability.  Even more generally,
I find that in the first $\sim1-10$\,Myr of evolution that the effective 
temperature of clusters and their embedded disks should be large than 
many tens of Kelvin, and for this reason  studies of protoplanetary
disk evolution for a much wider range of irradiation than has previously 
been considered should be carried out.

In Section \ref{section:temp}, I estimate the temperature of embedded star clusters, and in  
Section \ref{section:embed} I estimate their protoplanetary disk temperatures.  I then 
compare the relative importance of 
cluster irradiation and (1) host star irradiation in passive disks, and (2)
accretion power in active disks.
Generically, I find that cluster irradiation dominates the thermodynamics 
and structure of passive and active disks outside a semi-major axis of $\sim1-5$\,AU
around Solar-type stars.
In Section \ref{section:formation}, I discuss
planet formation via core accretion and gravitational instabilty, focusing
on the physics of the ice-line
and the $Q$-criterion.  I find that for star clusters above a stellar surface
density of $\sim 5\times10^3$\,\,M$_\odot$ pc$^{-2}$ there is no ice-line and that,
for even modest surface density clusters, the  $Q$-criterion indicates that 
only protoplanetary disks with masses larger than $\sim0.2$\,M$_\odot$ can
be gravitationally unstable around Solar-type stars.

In Section \ref{section:obs}, I estimate
the temperature at formation for a variety of systems, including 
Galactic star clusters and globular clusters, 
super star clusters in the nearby starburst M82,
the central nuclear clusters of galaxies, and compact galaxies at high redshift. 
Several globular clusters are likely to have exceeded $T_{\rm Ice}$
in at least part of their volume during formation, including 47 Tuc.
I extend the discussion to continuously star-forming disks and discuss planet formation 
in the context of local and high-$z$ starbursts, rapidly star-forming galaxies, and AGN,
finding that above a {\it gas} surface density of $\sim 10^4$\,\,M$_\odot$ pc$^{-2}$ 
(or star formation surface density of $\sim10^{3}$\,M$_\odot$ yr$^{-1}$ kpc$^{-2}$)
the temperatures of rapidly star-forming galaxies should exceed or approach $T_{\rm Ice}$.
Applying these results to the compact passive high-$z$ galaxies, I
find that most of the systems likely exceeded $T_{\rm Ice}$ during formation.

In Section \ref{section:discussion},
I discuss the results, focusing first on uncertainties in the estimates, including 
variations in the IMF, radiation transport with 
distributed sources,  estimates of the optical depth in star clusters, and 
the time dependence of cluster and planet formation.  With these uncertainties in mind, 
I discuss local globular clusters, and in particular 
suggest other transit/RV searches for hot Jupiters in Galactic globular clusters 
than have central temperatures below $T_{\rm Ice}$, but metallicities comparable 
to 47 Tuc. I provide a sketch of a calculation of the fraction of all star formation in 
the universe that occurred above a given stellar surface density and metallicity,
with implications for the $z=0$ planetary mass density.  Section \ref{section:conclusion}
provides a conclusion.

\begin{figure}
\centerline{\includegraphics[width=9cm]{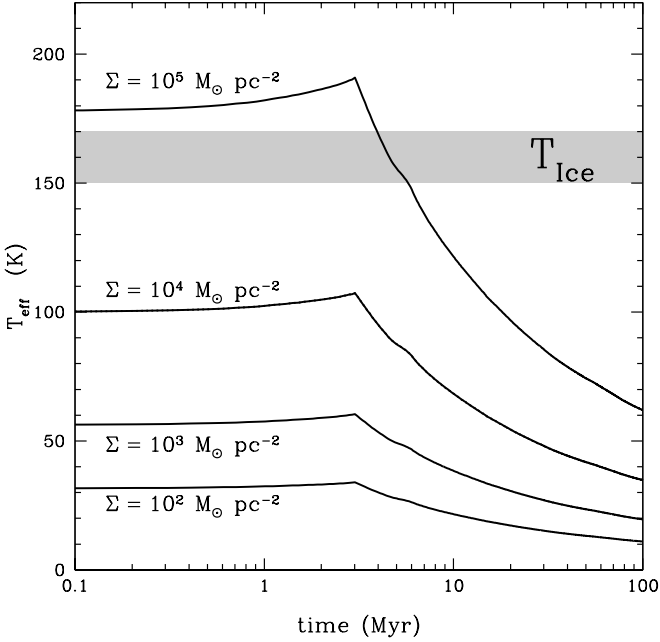}}
\caption{Effective temperature of star clusters as a function of time
for an instantaneous burst of star formation for $\Sigma=10^{5}-10^{2}$\,M$_\odot$ pc$^{-2}$ 
(top to bottom). The gray band shows $T_{\rm Ice}\simeq150-170$\,K.
After $t\simeq3$\,Myr, $T_{\rm eff}\propto t^{-1/3}$ with the normalization set by eq.~(\ref{teff}). }
\label{figure:l}
\end{figure}

\section{The Temperature of Star Clusters}
\label{section:temp}

A star cluster has mass $M$, radius $R$,
and a stellar surface density $\Sigma=M/4\pi R^2$.\footnote{We
define $\Sigma$ with the "4" since we are interested in the total flux, as seen from 
within the cluster.} 
If the stars form in a time less than 3\,Myr,\footnote{The dynamical timescale is 
$\sim10^5$\,yr\,\,$(R/{\rm pc}/\Sigma_4)^{1/2}$.} then the flux at 
formation can be estimated as 
\beq
F\approx\Psi\Sigma
\simeq1.6\times10^{7}\,{\rm L_\odot\,\,pc^{-2}}\,\Psi_{1500}\,\Sigma_{\,4},
\label{fstar}
\eeq
where $\Sigma_{x}=\Sigma/10^x$\,M$_\odot$ pc$^{-2}$ and 
$\Psi_{1500}=(\Psi/1500)$\,L$_\odot$/M$_\odot$ is the light-to-mass
ratio for a zero-age main sequence (ZAMS) stellar population
assuming a fully-populated standard IMF (e.g., Leitherer et al.~1999).\footnote{For 
clusters with small total mass $\lesssim10^3$\,M$_\odot$, the assumption
of a fully populated IMF breaks down and $\Psi$ is decreased.}
This flux corresponds to an effective temperature of 
\beq
T_{\rm eff}\simeq101\,{\rm K}\,\,\,(\Psi_{1500}\,\Sigma_{\,4})^{1/4}.
\label{teff}
\eeq
Figure \ref{figure:l} shows $T_{\rm eff}$ as a function of time for an 
instantaneous burst of star formation, using a Salpeter IMF from $1-100$\,M$_\odot$, 
for several values of $\Sigma$.   After $3-4$\,Myr, and over the first $100$\,Myr
the time-dependence of $T_{\rm eff}(t)$ is well approximated by $t^{-1/3}$.
The gray band shows the temperature for water ice sublimation,
$T_{\rm Ice}\simeq150-170$\,K.  The effect of variation in the IMF on $\Psi$ and 
$T_{\rm eff}$ is discussed in Section \ref{section:uncertain}.

Since the focus here is on the application to protoplanetary disks, one can ask about 
the total flux absorbed by each side of an optically-thick disk situated at the center
of an optically-thin spherically-symmetric star cluster with radial emissivity 
profile $j(r)=\Psi \rho$.  For illustration, taking a Hubble profile for the 
stellar density distribution,
$\rho(r)=\rho_0/[1+(r/r_0)^2]^{3/2}$, one obtains
$F(r=0)=\Psi\rho_0 r_0/4=\sigma_{\rm SB}T_{\rm eff}^4$, implying
\beq
T_{\rm eff}\simeq108\,{\rm K}\,\,\,(\Psi_{1500}\,\rho_{0,\,5}\,r_{0,\,{\rm pc}})^{1/4},
\eeq
where $r_{0,\,{\rm pc}}=r_0/{\rm pc}$ and $\rho_{0,\,5}=\rho_0/10^5$\,M$_\odot$ pc$^{-3}$
so that the equivalent stellar surface density inside $r_0$ 
is $\Sigma=M(r<r_0)/4\pi r_0^2\simeq1.7\times10^4$\,M$_\odot$ pc$^{-2}$.
More generally, the total energy density at any radius $r$ within an 
optically-thin cluster of emissivity $j(r)=\Psi\rho(r)$ is 
\beq
u(r)=\frac{1}{2c}\int_0^\infty j(r^{\,\prime})\left(\frac{r^{\,\prime}}{r}\right)
\ln\left|\frac{1+r^{\,\prime}/r}{1-r^{\,\prime}/r}\right|\,dr^{\,\prime}.
\label{uthin}
\eeq

\subsection{Optically-Thick Clusters}
\label{section:thick}

If the star cluster is compact compared to the optically-thick embedding
dust/gas distribution, the temperature of the star 
cluster is larger than $T_{\rm eff}$ because the radiation
field is trapped and diffusive, as in a simple stellar atmosphere.
In this case, the cluster temperature is
\beq
T_c^4\simeq\frac{3}{4}\left(\tau_R+\frac{2}{3}\right) T^4_{\rm eff},
\label{t}
\eeq
where $\tau_R$ is the Rosseland-mean optical depth through the 
surrounding dusty medium.
For the purposes of analytic estimates, 
in the temperature regime $160$\,K\,$\lesssim T \lesssim 1500$\,K, 
I approximate the Rosseland-mean dust opacity as constant:
\begin{equation}
\kappa_R\sim5 f_{\rm dg,\,MW}\,{\rm cm^2\,\, g^{-1}},
\label{kappahot}
\end{equation}
where I assume a Galactic dust-to-gas ratio ($f_{\rm dg,\,MW}=1$).
Comparing with Adams \& Shu (1986), Bell \& Lin (1994), and Semenov et al.~(2003),
this approximation agrees with more detailed calculations to a factor of $\sim2$
over this temperature range.
For lower temperatures ($T\lesssim160$\,K),
\beq
\kappa_R\sim\kappa_0 f_{\rm dg,\,MW} T^2,
\label{kappacool}
\eeq
where $\kappa_0\simeq2.4\times10^{-4}$\,cm$^2$ g$^{-1}$ K$^{-2}$ (Bell \& Lin 1994; Semenov et al.~2003).

Approximating the cluster as a geometrically thin shell of gas surface density
$\Sigma_g$ and gas fraction $f_g=\Sigma_g/\Sigma$, the total optical depth is
\beq
\tau_R\sim\kappa_R f_g\Sigma\sim2\,\kappa_R\,f_g\Sigma_{\,4}.
\label{tau}
\eeq
For $T_c\gtrsim160$\,K and $\tau_R\gg2/3$, the cluster temperature is then approximately
\beq
T_c\sim170\,{\rm K}\,\,(f_{\rm dg,\,MW}\,f_g\,\Psi_{1500})^{1/4}
\,\Sigma_{\,4}^{1/2},
\label{tchot}
\eeq
while for lower temperatures ($T\lesssim160\,{\rm K}$),\footnote{There is only a
small dynamic range in $f_g\Sigma$ where $\tau_R>2/3$ and $T_c<160$\,K. See Figure \ref{figure:temp}.}
\beq
T_c
\sim130\,{\rm K}\,(f_{\rm dg,\,MW}f_g\Psi_{1500})^{1/2}\,\Sigma_{\,3.8}.
\label{tccool}
\eeq
The radiation energy density is $u=a_r T_c^4$, where 
$a_r$ is the radiation constant.  These estimates for hot, high surface density 
embedded clusters are similar to those derived by Murray (2009).

\subsection{Optically-Thin Clusters}
\label{section:thin}

When the cluster is optically-thick to the UV/optical radiation from the stars, but optically-thin
to the re-radiated FIR emission, the temperature is $T^4\sim T_{\rm eff}^4/4\tau_R$,
and the energy density is given by equation (\ref{uthin}).
The critical surface density below which the cluster 
has $\tau_R<2/3$ can be derived roughly using equations (\ref{t}) and 
(\ref{kappacool}) as $\Sigma_{\tau_R\,=\,2/3}\simeq1500\,\,{\rm M_\odot 
\,\,pc^{-2}}\,(f_{\rm dg,\,MW}f_g)^{-2/3}\Psi_{1500}^{-1/3}$.

\subsection{UV \& Optically Optically-Thin Clusters}
\label{section:uvthin}

The cluster is optically-thin
to both the UV/optical radiation from massive stars and to the re-radiated FIR emission
when $\Sigma\lesssim5$\,M$_\odot$ pc$^{-2}$\,\,$(f_{\rm dg,\,MW}f_g)^{-1}$,
which may occur
after the cluster has evolved out of the embedded phase.  In this limit, $T_{\rm eff}$ is 
still given by equation (\ref{teff}) and  $u(r)$ is given by equation (\ref{uthin}), 
but the temperature of the radiation field is of course much hotter ($\sim10^4$\,K) 
than in the limits of Sections \ref{section:thick} and \ref{section:thin}.
This hard radiation field will super-heat dust grains at the surface of 
protoplanetary disks to a temperature larger than $T_{\rm eff}$ because of 
the emissivity properties of grains in the UV and FIR.

\begin{figure}
\centerline{\includegraphics[width=9cm]{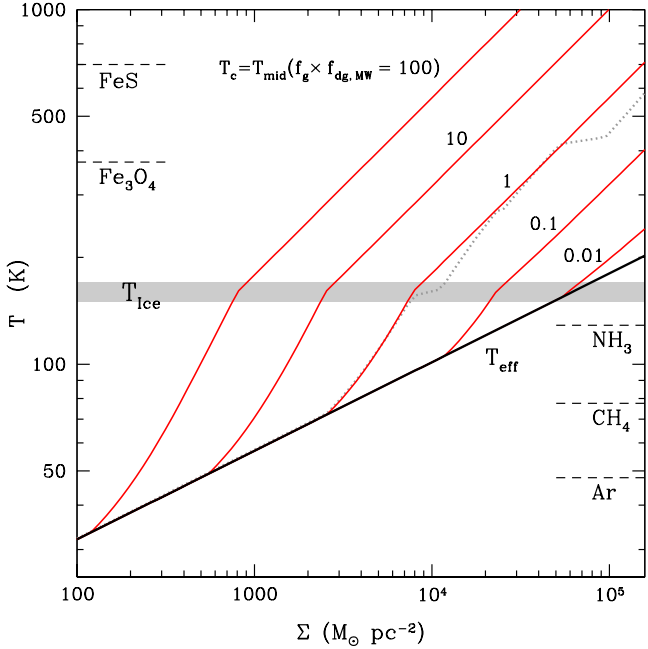}}
\caption{Temperature as a function of cluster stellar surface density.
The thick solid black line is $T_{\rm eff}$ for the cluster (eq.~\ref{teff})
for $t\lesssim3-4$\,Myr (see Fig.~\ref{figure:l}).
The red lines show the central cluster temperature $T_c$ for 
$f_{\rm dg,\,MW}\times f_g=100$, 10, 1, 0.1, and 0.01 (eq.~\ref{t}), using the 
cluster optical depth, $\tau_R$, and using the approximate opacities of 
eqs.~\ref{kappahot} and \ref{kappacool}.   A numerical solution to eq.~(\ref{t})
using the full opacity tables of Semenov et al.~(2003) for $f_{\rm dg,\,MW}=f_g=1$
is shown by the dotted line.  A protoplanetary disk
that is optically-thick to the cluster's FIR emission at temperature $T_c$ attains
a minimum temperature $T_{\rm mid}\sim T_c$ if the cluster has $\tau_R>1$.
For $\tau_R<1$, a  protoplanetary disk
that is optically-thick to the cluster's FIR emission at temperature $T_c$ attains
a minimum temperature $T_{\rm mid}\sim T_{\rm eff}$.
The gray band shows $T_{\rm Ice}\approx150-170$\,K.  
The horizontal dashed lines show condensation temperatures for troilite, magnetite,
ammonia, methane, and argon ices (highest to lowest; Lodders 2003).}
\label{figure:temp}
\end{figure}

\section{The Temperature of Protoplanetary Disks}
\label{section:embed}

Given the ambient radiation field in the cluster, one can estimate its importance
for the thermal structure of protoplanetary disks around individual stars within the cluster.
and compare the importance of cluster irradiation to the insolation from the 
host star and to the thermal structure determined by accretion in active disks. 

\subsection{Embedded Disks}

There are three primary regimes of interest.  To illustrate these, I adopt a simple 
model of a protoplanetary disk around a star of mass $M_\star$, radius $R_\star$, temperature
$T_\star$.  The disk has gas surface
density $\Sigma=\Sigma_0({\rm AU}/a)^x$,\footnote{For a 
minimum mass Solar nebula, $\Sigma_0\simeq10^3$\,g cm$^{-2}$ and $x\simeq3/2$
(Weidenschilling 1977).}
where $a$ is the semi-major axis, disk midplane 
temperature $T_{\rm mid}$  set by radiative and hydrostatic equilibrium, and vertical 
optical depth $\tau_V=\kappa_R(T_{\rm mid})\Sigma/2$, where $\kappa_R$ is 
given by equations (\ref{kappahot}) and (\ref{kappacool}) if the grain size distribution
and composition within the disk is the same as within the star cluster and if we 
consider only temperatures below the dust sublimation temperature at $\sim1500$\,K.\footnote{Note
that the grain size distribution of proto-planetary disks is likely different from that of the 
ISM, with a larger maximum grain size and different slope for the power-law size distribution
(e.g., D'Alessio et al.~1999, 2001).  These differences quantitatively 
affect the dividing lines between the different semi-major axis regimes discussed below,
but all qualitative conclusions remain unchanged.  More discussion of this issue
is provided in Sections \ref{section:active} and \ref{section:more_complete}.}

Neglecting all other sources of disk heating, including that from the central star and accretion,
the three regimes of cluster irradiation for the disk 
are as follows. At small $a$ the disk is 
optically-thick to radiation of temperature $T_c$, and to its own radiation $T_{\rm mid}$.
Here, $T_{\rm mid}\sim (u/a_r)^{1/4}$. 
Next,  at an intermediate range of $a$, the disk may be optically-thick to radiation of  
temperature $T_c$, but optically-thin to its own radiation $T_{\rm mid}$ and 
$T_{\rm mid}\sim (u/a_r\tau_V)^{1/4}$.
Finally, at large $a$, the disk is optically-thin to radiation of temperature $T_c$ and to 
its own radiation $T_{\rm mid}$ and 
$T_{\rm mid}\sim[\epsilon(T_c)/\epsilon(T_{\rm mid})]^{1/4}(u/a_r)^{1/4}$, where 
$\epsilon$ is the dust emissivity at temperature $T$.\footnote{Note that $\epsilon(T_c)/\epsilon(T_{\rm mid})$ 
would only be much different than unity in the case of a 
UV/optically optically-thin cluster as discussed in Section \ref{section:uvthin}.}

These limits show that a protoplanetary disk embedded in a cluster of 
effective temperature $T_{\rm eff}$ (eq.~\ref{teff}) has $\sim T_{\rm eff}$ as its minimum
temperature.  If the cluster is optically-thick, then the minimum temperature of the optically-thick
regions of the disk is $\sim T_c$ (eqs.~\ref{tchot} \& \ref{tccool}), which exceeds $T_{\rm eff}$
by a factor of $\sim\tau^{1/4}$ (eq.~\ref{t}).  

Figure \ref{figure:temp} shows several temperatures of interest as a function
of cluster stellar surface density for a newly-born cluster ($t\lesssim3-4$\,Myr).  
The thick solid black line shows $T_{\rm eff}$ (eq.~\ref{teff})
for $\Psi=1500$\,L$_\odot$/M$_\odot$ (see Fig.~\ref{figure:l}).  The red
solid lines show $T_c$ for $\tau_R>2/3$ obtained by solving equation (\ref{t}), using the 
simplified opacities of equations (\ref{kappahot}) and (\ref{kappacool}), and for 
$f_{\rm dg,\,MW}\times f_g=100$, 10, 1, 0.1, and 0.01. The break in each red line occurs 
at $T_c=160$\,K at the break in 
the assumed opacity law (eqs.~\ref{tchot} \& \ref{tccool}).
A more complete treatment of 
the Rosseland-mean opacity, as presented in Bell \& Lin (1994) or Semenov et al.~(2003),
introduces inflections in $T_c(\Sigma)$ at each break in the opacity curve, as shown 
in the dotted curve, which gives the full solution to  equation (\ref{t}) using the
opacity tables of Semenov et al.~(2003) for $f_{\rm dg,\,MW}\times f_g=1$. The close
correspondence between the dotted curve and the red curve validates our use of 
equations (\ref{kappahot}) and (\ref{kappacool}) in making analytic estimates below.
The gray band shows the temperature for water 
ice sublimation $T_{\rm Ice}\simeq150-170$\,K.
The short horizontal 
dashed lines show condensation temperatures for troilite ($\simeq700$\,K) and magnetite ($\simeq370$\,K), 
and then the ammonia, methane, and argon ices (Lodders 2003).
For times $\gtrsim3-4$\,Myr, $T_{\rm eff}$ and $T_c$ decrease approximately as $t^{-1/3}$
(Fig.~\ref{figure:l}).

\subsection{Cluster Irradiation versus Host Star Irradiation}
\label{section:host}

It is useful to compare the incoming flux from the combined stellar population of the star cluster 
with the flux from the central star in a passive flared disk (Chiang \& Goldreich 1997; hereafter CG97):
\beq
T_{\rm eff,\,\star}\approx
\left(\frac{\alpha R_\star^2}{4a^2}\right)^{1/4}
\hspace{-.2cm}T_\star\approx61\,{\rm K}\,\,a_{\rm 10\,AU}^{-3/7}\,\,R_{\odot}^{1/2}\,T_{\odot},
\label{teffstar}
\eeq
where $\alpha\approx0.005a_{\rm AU}^{-1}+0.05a_{\rm AU}^{2/7}$ is the flaring angle, 
$T_{\odot}=T_\star/6000$\,K, $R_{\odot}=R_\star/$\,R$_\odot$, 
and where the 
second approximate equality follows from ignoring the first term in $\alpha$ at large $a$.  

As discussed in CG97, the 
hot stellar irradiation produces a superheated dust layer (of temperature $T_{\rm dust}$) 
at the surface of the disk, which re-radiates
approximately half of the absorbed light back into the disk, maintaining $T_{\rm mid}$.
As in the case of cluster irradiation alone, there are three regimes of irradiation
from the central star separated by semi-major axis: (1) at small $a$ ($\lesssim100$\,AU for CG97 parameters), 
the disk is optically-thick to both $T_{\rm dust}$ and $T_{\rm mid}$; 
(2) at an intermediate range of $a$ the disk is optically-thick to $T_{\rm dust}$, 
but optically-thin to radiation of temperature $T_{\rm mid}$; (3) at large
$a$ ($\gtrsim200$\,AU in CG97) the disk is optically-thin to both.  In all cases, 
$T_{\rm mid}$ for the passive disk is set
by the incoming flux from the central star, $T_{\rm eff,\,\star}$ in equation (\ref{teffstar}),
but with an overall correction for the fact that only half of the incoming flux is radiated into the disk, 
a correction for the vertical disk optical depth in regime (2), and a correction
for the ratio of emissivities between $T_{\rm dust}$ and $T_{\rm mid}$ in regime (3).

We can now compare the importance of cluster irradiation to that from the 
central star directly. When the cluster is optically-thick to its own FIR radiation ($\tau_R>2/3$),
the relevant comparison is between $T_c$ in equations (\ref{tchot}) and (\ref{tccool}) 
and $T_{\rm eff,\,\star}$. Setting  $T_c=T_{\rm eff,\,\star}$, one derives the critical 
semi-major axis outside of which the cluster radiation field dominates the radiation
field from the central star: 
\beq
a_{\rm crit}\sim0.9\,\,{\rm AU}
\left[\frac{R^2_{\odot}T_{\odot}^{4}}{f_{\rm dg,\,MW}\,f_g\,\Psi_{1500}\Sigma^2_4}\right]^{7/12}
(T_c>160{\rm \,K})
\label{acritirrhot}
\eeq
\beq
a_{\rm crit}\sim0.6\,\,{\rm AU}
\left[\frac{R_{\odot}T_{\odot}^{2}}{f_{\rm dg,\,MW}\,f_g\,\Psi_{1500}\Sigma^2_{4}}\right]^{7/6}
(T_c<160{\rm \,K})
\label{acritirrcold}
\eeq
When the cluster is optically-thin to its own FIR radiation the relevant comparison is between
$T_{\rm eff}$ from the cluster (eq.~\ref{teff}) and $T_{\rm eff,\,\star}$:
\beq
a_{\rm crit}\sim3.0\,\,{\rm AU}
\left[\frac{R^2_{\odot}T_{\odot}^{4}}{\Psi_{1500}\Sigma_4}\right]^{7/12}
(\tau_R<2/3).
\label{acritirreff}
\eeq
Since at these small values of the semi-major axis the disk is optically-thick to the 
incoming FIR radiation from the cluster, the re-radiated emission from the hot surface 
disk dust layer, and to the re-radiated emission from the disk midplane, these comparisons 
show that the cluster radiation field dominates the temperature and structure of passive 
irradiated disks for $a\gtrsim1$\,AU around typical
stars in a cluster with $\Sigma=10^4$\,M$_\odot$ pc$^{-2}$; for $a\lesssim a_{\rm crit}$,
$T_{\rm mid}$ is set by stellar irradiation, but for $a\gtrsim a_{\rm crit}$, the cluster sets 
$T_{\rm mid}$.  Its minimum value is $T_{\rm eff}$ (black solid line
in Fig.~\ref{figure:temp}) in the case of an optically-thin cluster, and it reaches  $T_c$ in the 
case of a cluster that is optically-thick to its own FIR radiation.  

It is worth noting that because the geometry of cluster irradiation is different than for
the central star, the vertical structure of the disk will be modified.
In particular, the flaring angle in a passive disk irradiated only by a central star is expected
to scale as $h/a\propto a^{2/7}$ (CG97), whereas for a passive isothermal disk irradiated 
isotropically by the cluster  one expects $h/a\propto a^{1/2}$.  This stronger
flaring will increase the insolation from the host star, and in a self-consistent model for the 
disk, cluster, and star, and one expects the estimates above for $a_{\rm crit}$ to be modified
quantitatively.

\begin{figure}
\centerline{\includegraphics[width=9cm]{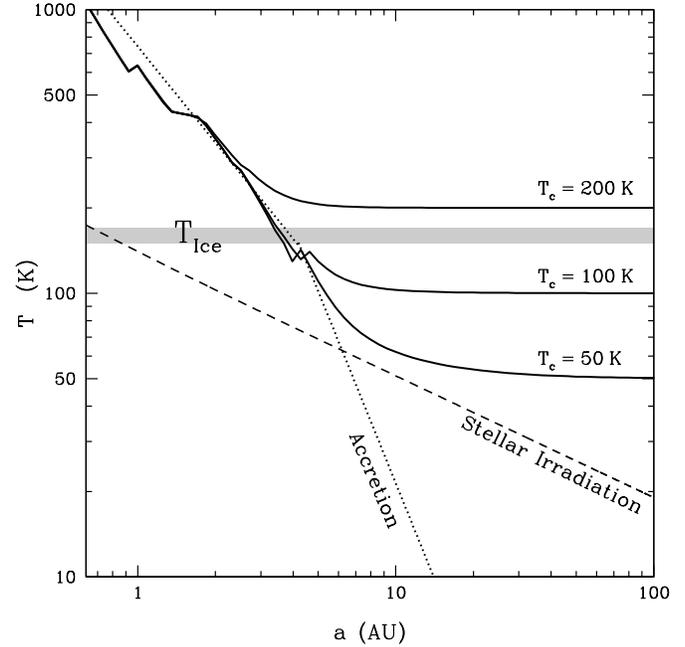}}
\caption{Protoplanetary disk temperature as a function of semi-major axis $a$ 
for a star with $M_\star=$\,M$_\odot$, $T_\star=T_\odot=6000$\,K, $R_\star=$\,R$_\odot$,
and a disk with $\Sigma=3\times10^3$\,g cm$^{-2}\,\,({\rm AU}/a)^{3/2}$.
The gray band shows $T_{\rm Ice}=150-170$\,K, the 
short dashed black line shows $T_{\rm eff,\,\star}/2^{1/4}$ (eq.~\ref{teffstar}),
which approximates the contribution to the midplane temperature from stellar irradiation alone.
The dotted line shows the analytic estimates of $T_{\rm mid,\,acc}$ assuming an accretion rate 
of $10^{-8}$\,M$_\odot$ yr$^{-1}$ using eqs.~(\ref{tmidacchot}) and (\ref{tmidacccool}).  The heavy black lines
show the calculated value of midplane temperature of the disk as a function of $a$, taking into
account host star insolation, accretion, and cluster irradiation using the 
opacities Semenov et al.~(2003) with 
$T_c=50$, 100, and 200\,K, respectively, from lowest to highest. Higher $T_c$ in general
moves the ice-line to larger $a$, and when $T_c>T_{\rm Ice}$ (e.g., for $T_c=200$\,K), 
the ice-line disappears entirely.}
\label{figure:cg}
\end{figure}

\subsection{Cluster Irradiation versus Accretion}
\label{section:active}

For an active disk with steady accretion rate $\dot{M}_{-8}=\dot{M}/10^{-8}$\,M$_\odot$ yr$^{-1}$
around a star of mass $M_\star$ measured in units of $M_\odot$,
the flux from the disk is 
\beq 
T_{\rm eff,\, acc}=\left[\frac{3}{8\pi}\frac{\dot{M}\Omega^2}{\sigma_{\rm SB}}\right]^{1/4}
\approx26\,{\rm K}\,\,(\dot{M}_{-8}M_\star /a_{\rm \,5\,AU}^{3})^{1/4}.
\eeq
Ignoring other sources of heating, the midplane temperature of the disk is larger 
than $T_{\rm eff,\,acc}$ by a factor of $[(3/4)\tau_{V}]^{1/4}$.
Taking values characteristic of a minimum mass Solar nebula for illustration ---
$\Sigma(a)=\Sigma_0({\rm AU}/a)^{3/2}$ with $\Sigma_{0,3}=\Sigma_0/10^3$\,g cm$^{-2}$ --- 
the midplane temperature is 
\beq
T_{\rm mid,\,acc}\approx260\,{\rm K}\,(\Sigma_{0,3}\dot{M}_{-8}M_\star)^{1/4}\,a_{\rm 2\,AU}^{-9/8} 
\,\,\,(T_{\rm mid}>160{\rm\, K}) 
\label{tmidacchot}
\eeq
\beq
T_{\rm mid,\,acc}\approx59\,{\rm K}\,(\Sigma_{0,3}\dot{M}_{-8}M_\star)^{1/2}\,a_{\rm 5\,AU}^{-9/4} 
\,\,\,(T_{\rm mid}<160{\rm\, K}),
\label{tmidacccool}
\eeq
where the top expression assumes equation (\ref{kappahot}) and the lower expression 
assumes equation (\ref{kappacool}).\footnote{For 
a proto-planetary disk dust grain size distribution, the Rosseland-mean opacity is decreased
by a factor of $\sim2-3$ relative to the estimates of equations (\ref{kappahot}) and  (\ref{kappacool})
in the temperature range of interest (D'Alessio et al.~1999, 2001).  In general, this lowers the 
midplane temperature of active disks with respect to the estimates in equations (\ref{tmidacchot}) 
and (\ref{tmidacccool}).  This moves the ice-line in active disks to smaller semi-major axes 
(Sasselov \& Lecar 2000; Lecar et al.~2006).  Setting equation (\ref{tmidacchot}) equal to $T_{\rm Ice}=170$\,K
one  finds that $a_{\rm Ice}\simeq2.8\,{\rm AU}[(\kappa_R/4\,{\rm cm^2/g})\Sigma_{0,3}\dot{M}_{-8}M]^{8/9}$,
whereas a calculation with the D'Alessio et al.~(2001) opacities by Lecar et al.~(2006)
finds $a_{\rm Ice}\simeq1.7$\,AU for the same parameters.}  
The above scalings show that accretion heating generally
dominates $T_{\rm mid}$ at $a\lesssim{\rm few}\,$AU, whereas passive heating from the central star
dominates $T_{\rm mid}$ at larger semi-major axis (CG97; compare eqs.~\ref{tmidacccool} and \ref{teffstar}).

In order to evaluate the importance of cluster irradiation for the structure of active
disks, the relevant comparison is then between $T_{\rm mid,\,acc}$ and either $T_c$ in the 
case of a FIR optically-thick cluster or $T_{\rm eff}$ in the case of a FIR optically-thin
cluster.  Roughly, $T_{\rm mid}$ will only be strongly modified when 
$\max[1,\tau_R]\Psi\Sigma$ for the cluster exceeds $\tau_V\dot{M}\Omega^2/8$ for the optically-thick
regions of the protoplanetary disk.  
For optically-thick clusters with $\tau_R>2/3$ the critical semi-major axis
beyond which cluster irradiation dominates accretion heating in setting $T_{\rm mid}$ is
\beq
\hspace*{-.2cm}a_{\rm crit}\sim2.9\,\,{\rm AU}
\left[\frac{\Sigma_{0,3}\dot{M}_{-8}M_\star}{f_{\rm dg,\,MW}\,f_g\,\Psi_{1500}\Sigma^2_4}\right]^{2/9},
\label{acritaccthick}
\eeq
which is correct for either $T_c,\,T_{\rm mid}>160{\rm \,K}$ or $T_c,\,T_{\rm mid}<160{\rm \,K}$.
When the cluster is optically-thin ($\tau_R<2/3$), equations (\ref{teff}) and 
(\ref{tmidacccool}) imply that
\beq
a_{\rm crit}\sim3.9\,\,{\rm AU}
\left[\frac{(\Sigma_{0,3}\dot{M}_{-8}M_\star)^2}{\Psi_{1500}\Sigma_{4}}\right]^{1/9}
(\tau_R<2/3).
\label{acritaccthin}
\eeq

These comparisons show that the cluster radiation field dominates the disk structure
of active accreting disks for semi-major axes larger than  
$\sim{\rm few}-10$\,AU around a typical star with an active accretion disk
embedded in a cluster with $\Sigma\gtrsim10^2$\,M$_\odot$ pc$^{-2}$.  The weak
dependencies of $a_{\rm crit}$ on $\Sigma$ and $\dot{M}$ in equation (\ref{acritaccthin}) 
guarantee that beyond a $\sim{\rm few}-10$\,AU, cluster irradiation dominates
accretion, even for accretion rates as high as $5\times10^{-7}$\,M$_\odot$ yr$^{-1}$
and cluster surface density $\Sigma$ as low as $10^{2.5}$\,M$_\odot$ pc$^{-2}$.

However, as I discuss in Sections \ref{section:grav} and \ref{section:accretion},
the cluster-irradiated massive disks most prone to gravitational instability will have accretion
rates as much as $\sim10^2-10^4$ times higher than used in the estimates above, and particularly
at early times.  The normalization of equations (\ref{acritaccthick}) and 
(\ref{acritaccthin}) increases to $\simeq22$ and $\simeq11$\,AU, respectively, 
for $\dot{M}=10^{-4}$\,M$_\odot$ yr$^{-1}$.
Additionally, the accretion luminosity in such a system would 
dominate the passive heating contribution in Section \ref{section:host},
pushing the critical semi-major axis at which the cluster irradiation dominates the host 
star to much larger values than given in equations (\ref{acritirrhot})-(\ref{acritirreff}).
Most importantly, such high accretion rates would significantly decrease the disk lifetime.

\subsection{A More Complete Calculation}
\label{section:more_complete}

In order to make these comparisons between cluster irradiation and host star irradiation 
and accretion explicit, 
in Figure \ref{figure:cg} I show the midplane temperature of a protoplanetary
disk (thick solid lines), including all three effects, for $T_c=200$, 100, 
and 50\,K (highest to lowest). 
 
The dashed line (``stellar irradiation'') shows $T_{\rm eff,\,\star}$ (eq.~\ref{teffstar}; CG97),
and the dotted line (``accretion'') shows the analytic approximations for $T_{\rm mid,\,acc}$ in 
equations (\ref{tmidacchot}) and (\ref{tmidacccool}). 
The midplane temperature is calculated by solving the implicit equation (see, e.g., Sirko \& Goodman 2003)
\beq
T_{\rm mid}^4=\frac{3}{4}\left[\tau_{V}+\frac{4}{3}+\frac{2}{3\tau_{V}}\right]T_{\rm eff,\,acc}^4
+\left(1+\frac{1}{\tau_{V}}\right)(T_{\rm eff,\,\star}^4+T_c^4),
\eeq
where $\tau_V(T_{\rm mid})$ is the vertical optical depth to the disk's own
radiation, and where I have assumed the opacities of Semenov et al.~(2003).  As in CG97,
the last term allows for the fact that the disk may be optically-thick to the incoming 
radiation field, but optically-thin to its own re-radiated emission.

The calculation is not self-consistent in the sense that it assumes $\dot{M}$
is constant throughout the disk, and that $T_{\rm eff,\,\star}$ is given 
by equation (\ref{teffstar}).  In reality, the disk structure, 
and hence the amount of stellar irradiation as a function of $a$, should deviate from 
equation (\ref{teffstar}) because of accretion, and because
the disk becomes isothermal ($T_{\rm mid}\rightarrow T_c$)
at large $a$.  In addition, because the scale height is determined without
reference to $\dot{M}$, in such a disk
as plotted in Figure \ref{figure:cg}, the viscosity parameter $\alpha$ (not to 
be confused with $\alpha$ in eq.~\ref{teffstar}) would 
continuously vary.  For these reasons, the calculations shown in  Figure \ref{figure:cg}
are meant to be a sketch.
Even so, Figure \ref{figure:cg} dramatically illustrates
the importance of cluster irradiation.  Even for $T_c=50$\,K, the thermal 
structure of the disk is completely modified from the expectation of 
passive stellar irradiation for semi-major axes  larger than $\sim10$\,AU.
Comparing $T_c=50$\,K to $T_c=100$\,K, one sees that the ice line moves out
slightly in $a$ and that the disk structure is modified from the expectation
of the simply active disk beyond $a\gtrsim 4$\,AU.  For $T_c=200$\,K, the ice line
completely disappears, and the structure of the disk changes noticeably 
beyond $\gtrsim2$\,AU. 

Detailed self-consistent calculations of active disks with both cluster and host star irradiation 
are left to a future effort.

\subsection{Other Regimes}

Many more regimes than those listed in equations (\ref{acritirrhot})-(\ref{acritirreff})
and (\ref{acritaccthick})-(\ref{acritaccthin}), and represented in Figure \ref{figure:cg},
could be enumerated. For example, 
when the cluster is
UV/optically optically-thin and the cluster has the hard radiation field 
characteristic of a young stellar population ($\sim10^4$\,K, Section \ref{section:uvthin}) 
an active and irradiated protoplanetary disk will be optically-thick to this incoming UV radiation at
much larger semi-major axis than the FIR radiation of an optically-thick cluster, thus effecting
at what semi-major axis the cluster, the accretion, or the insolation from 
the host star dominates the disk structure and thermodynamics. In addition, such a cluster
radiation field would produce super-heated dust layers on all its protoplanetary disks,
thus affecting the dust-reprocessed radiation field the disks would see at larger $a$.

\section{Planet Formation}
\label{section:formation}

The estimates above show that the temperature within compact embedded
star clusters is expected to be very high.  Protoplanetary disks within
these clusters can be expected to have a minimum temperature of order 
$T_{\rm eff}$ given by equation (\ref{teff}), and with time evolution 
shown in Figure \ref{figure:l}.  Highly embedded clusters reach considerably 
higher central temperatures, as shown by the red lines in Figure 
\ref{figure:temp} for different values of the dust-to-gas ratio and 
cluster gas fraction.  As discussed in Section \ref{section:embed} and shown
in Figure \ref{figure:cg},
the radiation field from a young cluster is likely to dominate the 
thermodynamics and structure of both passive and accreting 
protoplanetary disks on scales larger than $\sim1-10$\,AU,
depending on the parameters of the system in consideration. 
Here, I consider the zeroth-order effects on planet formation 
in the core accretion and gravitational instability pictures for 
the formation of gas- and ice-giant planets.

\subsection{Core-Accretion}
\label{section:core}

The dominant picture for the formation of gas and ice giant planets is 
the core accretion model (Mizuno 1980; Pollack et al.~1996), which
requires a protoplanetary core to grow in the disk to a critical mass of 
$\sim10$\,M$_\oplus$, thereby initiating runaway accretion of gas from the disk and 
rapid growth to the isolation mass (e.g., Armitage 2010).  Critical to the core
accretion model is the ``ice line,'' the semi-major axis where the temperature of the 
disk drops below the sublimation temperature of water ice, because 
here the surface density of 
all condensables is $\sim3-5$ times higher than that in only refractories.
It is this ``extra'' material, which exists only outside the ice line, that
enables the formation of $\sim10$\,M$_\oplus$ mass cores and the core accretion 
phenomenon. In standard models of disk evolution including
irradiation from the central star and accretion, the ice line typically occurs
at $a\sim2-3$\,AU (e.g., Sasselov \& Lecar 2000; Lecar et al.~2006; Kennedy \& Kenyon 2008), and 
it is this fundamental scale that is posited to explain the dichotomy between the 
terrestrial and gas/ice giants in the solar system.

Looking at Figure \ref{figure:temp}, one finds that in star clusters with 
high surface densities and/or high gas fractions or dust-to-gas ratios,
$T_c$ is greater than the ice line temperature of $T_{\rm Ice}\simeq150-170$\,K
for a fully populated initial mass function, for times less than $\simeq3$\,Myr.
Setting $T_c=T_{\rm Ice}$ in equation (\ref{tchot}), one derives the critical 
stellar surface density above which $T_c > T_{\rm Ice}$ and there is no ice 
line in the star cluster is
\beq
\Sigma_{\rm Ice}\simeq6\times10^3\,\,{\rm M_\odot\,\,pc^{-2}}\,\,\left(\frac{T_{\rm Ice}}{150\,{\rm K}}\right)^2
(f_{\rm dg,\,MW}\,f_g\,\Psi_{1500})^{-1/2}.
\label{sigmaice}
\eeq
Note the metallicity dependence in the denominator of this expression, which 
shows that a high value of the dust-to-gas ratio drives the system to higher 
temperatures, which could completely inhibit giant planet formation in 
the core accretion theory.
As shown in Figure \ref{figure:temp} and equation (\ref{sigmaice}), 
for super-solar metallicities or larger gas fractions, $\Sigma_{\rm Ice}$ 
decreases: for $f_{\rm dg,\,MW}\times f_g=100$,
$\Sigma_{\rm Ice}\simeq4\times10^2$\,M$_\odot$ pc$^{-2}$.  In contrast, for much
lower $f_{\rm dg,\,MW}$, as might occur in metal-poor proto-globular clusters with 
$[{\rm Fe/H}]=-2$ so that $f_{\rm dg,\,MW}\times f_g=0.01$,
$\Sigma_{\rm Ice}$ increases to $\sim5\times10^{4}$\,M$_\odot$ pc$^{-2}$. 

If giant planets in fact form by core accretion, then because this is a critical phenomenon
requiring  a large amount of condensable material, one expects clusters 
born with $\Sigma>\Sigma_{\rm Ice}$ to have only terrestrial rocky planets, and for
those planets to be devoid of water.  Such a disk is sketched in Figure \ref{figure:cg},
which shows that for $T_c=200$\,K (top thick solid line), the ice line simply disappears. 
I discuss the search for giant planets in the relatively 
metal-rich globular cluster 47 Tuc ($[{\rm Fe/H}]=-0.76$), the discovery of 
a giant planet in NGC 6121, and other globular clusters 
in Sections \ref{section:obs} and \ref{section:discussion}.

\begin{figure}
\centerline{\includegraphics[width=9cm]{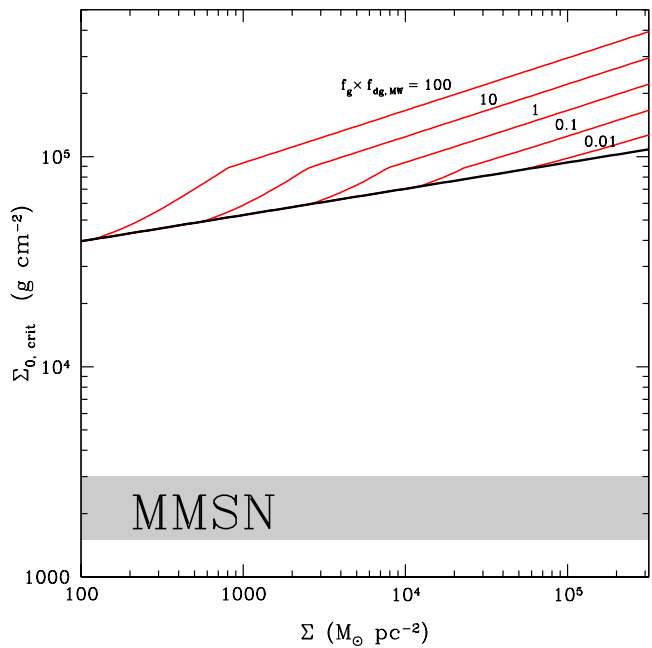}}
\caption{Critical protoplanetary surface density normalization
$\Sigma_{0,\rm \,crit}$ (see eqs.~\ref{grav1}-\ref{grav3}) above which the disk has $Q<1$ for 
$f_g\times f_{\rm dg,\,MW}=100-0.01$ (top to bottom; same as Fig.~\ref{figure:temp}),
assuming $\Sigma(a)=\Sigma_0({\rm AU}/a)^{3/2}$.  The normalization for the 
MMSN is shown for comparison as the gray band.  Very large $\Sigma_0$, and correspondingly large 
disk mass, is required for $Q<1$ in stellar clusters of essentially any surface density. }
\label{figure:tc}
\end{figure}

\subsection{Gravitational Instability}
\label{section:grav}

Gas- and ice-giant planets may also form by 
gravitational instability (Boss 1997, 2000; Pickett et al.~2003;  Durisen et al.~2007).  
Planet formation by gravitational instability puts two necessary 
physical requirements on the structure of the protoplanetary disk
that must be satisfied simultaneously: 
(1) Toomre's $Q$ must be less than $\sim1$ and (2) 
the cooling timescale must less than the local orbital
timescale (Gammie 2001).   Recent work has shown
that these requirements  can only be met at large 
semi-major axes around stars with very massive disks (Matzner \& Levin 2005; 
Rafikov 2005).  Such massive protoplanetary disks 
that become gravitationally unstable at large $a$ have also recently
been argued to produce low-mass binary stellar companions instead of 
planets (Boley 2009; Kratter et al.~2010; see also Bonnell \& Bate 1994, Bate 2000).  
 
Effects on $Q$ by irradiation
have been considered in the context of IR re-radiation by protostellar
envelopes by Matzner \& Levin (2005), and in the context of disk  models
by D'Alessio et al.~(1997), Cai et al.~(2008), Rafikov (2009), Vorobyov \& Basu (2010),
Rice et al.~(2011), Kratter \& Murray-Clay (2011), and  Zhu et al.~(2012).

The most obvious 
consequence of strong irradiation in the cluster environments 
investigated here is that Toomre's $Q$ is larger at fixed protoplanetary disk surface density
relative to a non-irradiated  protoplanetary disk of the same properties.  Thus, larger surface density 
is required to drive the disk gravitationally unstable.
Taking $c_s=\sqrt{k_B T/\mu}$ with $\mu=2.3m_p$, $\Omega=(GM/a^3)^{1/2}$, $M_\star=M_\star/{\rm M}_\odot$, 
and assuming that $\Sigma_g(a)=\Sigma_0(a_0/a)^{3/2}$, 
the critical temperature above which 
$Q=1=c_s\Omega/\pi G\Sigma_g$ scales as $T_{\rm Q=1}\propto\Sigma_0^2/M_\star$
with no $a$-dependence.
In this case,  the critical normalization $\Sigma_{0,\,{\rm crit}}$
above which the disk will have $Q<1$ is
\beq
\Sigma_{0,\,{\rm crit}}\simeq5.7\times10^4\,\,{\rm g\,\,cm^{-2}}\,\,M_\star^{1/2}
(\Psi_{1500}\Sigma_4)^{1/8}
\label{grav1}
\eeq
for an optically-thin cluster, and it increases to
\beq
\Sigma_{0,\,{\rm crit}}\simeq7.8\times10^4\,\,{\rm g\,\,cm^{-2}}\,\,M_\star^{1/2}
(f_{\rm dg,\,MW}f_g\Psi_{1500}\Sigma^2_4)^{1/8}
\label{grav2}
\eeq
and 
\beq
\Sigma_{0,\,{\rm crit}}\simeq1.1\times10^5\,\,{\rm g\,\,cm^{-2}}\,\,M_\star^{1/2}
(f_{\rm dg,\,MW}f_g\Psi_{1500}\Sigma^2_{4})^{1/4}
\label{grav3}
\eeq
for $T_c>$ and $T_c<160$\,K, respectively.  These values are $\sim50-100$
times the typical value assumed for the MMSN  ($\Sigma_0=1.5-3\times10^3$\,g cm$^{-2}$).
The implied disk masses are   $M_D\sim0.6-0.8$\,M$_\odot$ for an 
outer disk semi-major axis of 50\,AU.

Figure \ref{figure:tc} shows $\Sigma_{0,\,{\rm crit}}$ as a function of the cluster 
stellar surface density $\Sigma$ for the same parameters as in Figure \ref{figure:temp}:
$f_g\times f_{\rm dg,\,MW}=100$, 10, 1, 0.1, 0.01 (red lines, top to bottom; eqs.~\ref{grav2}, \ref{grav3}) 
and for $T=T_{\rm eff}$ (black line; eq.~\ref{grav1}).  The normalization for the
MMSN is shown as the gray band.  Thus, for disks embedded in star clusters with 
surface density $\Sigma$, and given $f_g\times f_{\rm dg,\,MW}$, the protoplanetary disks must 
have $\Sigma_0>\Sigma_{0,\,{\rm crit}}$ for $Q<1$.  This simple comparison
assumes that $T_{\rm mid}=T_c$, which is justified for large values of 
 $\Sigma_{0,\,{\rm crit}}$ required by equations (\ref{grav1})-(\ref{grav3})
since the vertical optical depth in the disk is larger than unity all the way
out to $\sim100$\,AU.

For different protoplanetary disk profiles, the scalings become more complicated.
For $\Sigma_g(a)=\Sigma_0(a_0/a)^{x}$, $x=1$, and $a_0={\rm AU}$,
the critical temperature above which  $Q=1$ is
\beq
T_{\rm Q=1}=62\,\,{\rm K}\,a_{20\rm AU}\,\Sigma_{0,\,4}^2/M_\star.
\label{tq1}
\eeq
I have purposefully scaled all quantities for a
massive disk, with total mass $M_D\simeq2\pi a_0 \Sigma_{0} \times20{\rm AU}\simeq0.14\Sigma_{0,\,4}$\,M$_\odot$.
Comparing with  Figures \ref{figure:l} and \ref{figure:temp} one
sees that for most of the parameter
regime considered $T_{\rm eff}$ for the cluster exceeds $T_{\rm Q=1}$ and that the 
disks will have $Q>1$ for $a<20$\,AU.  For the optically-thick
clusters with $T_c>T_{\rm eff}$, $T_c$ can in many cases dramatically exceed $T_{\rm Q=1}$,
as shown in Figure \ref{figure:temp}.

However, note that $T_{\rm Q=1}$ depends strongly on $\Sigma_0$, $a$, and $M_\star$.
For an isothermal disk with $x=1$, $Q\propto a^{-1/2}$, and thus equation (\ref{tq1})
can be interpreted as saying that in order for the disk to have a region with $Q<1$
in a cluster with temperature equal to $T_{\rm Q=1}$, the disk need only extend to a 
semi-major axis beyond 20\,AU (for the parameters of eq.~\ref{tq1}).  For example, 
equating $T_{\rm Q=1}$ with the
expression with the effective temperature of a star cluster of surface density $\Sigma$
in equation (\ref{teff}), one derives the critical semi-major axis of the protoplanetary disk
beyond which $Q<1$ in a disk with $x=1$:
\beq
a_{{\rm Q=1}}\simeq 33\,\,{\rm AU}\,\,M_\star\frac{(\Psi_{1500}\Sigma_4)^{1/4}}
{\Sigma_{0,\,4}^2\,\,a_{0,\,\rm AU}^2}.
\eeq
Alternatively, one can think of this as a limit on the disk mass $M_D$:
\beq
M_{D,\,{\rm Q=1}}\simeq 0.18\,\,{\rm M_\odot}\,\,\left(M_\star\,\,a_{20\,{\rm AU}}\right)^{1/2}
(\Psi_{1500}\Sigma_4)^{1/8},
\eeq
which says that for $M_D>M_{D,\,{\rm Q=1}}$ in a cluster of stellar surface density $\Sigma$,
the disk has $Q<1$ at $a\ge20$\,AU.
Thus, for gravitational instability to operate at, say, $a=100$\,AU in a disk with 
$x=1$ around a solar-mass star, in a 
cluster of nearly any stellar surface density $\Sigma$ (since $M_D\propto\Sigma^{1/8}$), the disk mass must
be very large: $M_D\simeq0.2-0.4$\,M$_\odot$.

The above estimates are only for optically-thin clusters.  For the 
case of an optically-thick cluster with $\tau_R>2/3$, the limit on $M_D$ increases,
but again depends on $x$.

Overall, these estimates imply that cluster irradiation strongly suppresses
gravitational instability by increasing $Q$ above its nominal value for a disk in isolation.
The additional requirement on the formation of planets by gravitational
instability is that the cooling time of the disk should be less than the 
characteristic orbital time.  The reader is referred to Kratter et al.~(2009),
Cai et al.~(2008), Rafikov (2009), Rice et al.~(2011), Kratter \& Murray-Clay (2011), and Zhu et al.~(2012)
 for a discussion of this criterion including irradiation.  I note here only 
that the magnitude of the cluster irradiation indicated by Figures (\ref{figure:l}),
and (\ref{figure:temp}), and which I estimate for observed stellar systems in 
Section \ref{section:obs}, in general dramatically exceeds what has been
so far considered in the literature for protoplanetary systems forming
by gravitational instability.

\subsection{Accretion \& Migration}
\label{section:accretion}

Because of the increase in sound speed at large semi-major axis in 
disks with strong cluster irradiation with respect to those without, one generically expects 
active disks to accrete at higher rates,  yielding a shorter overall planet formation 
timescale, and for planets
embedded within them to migrate faster, if the predominant migration 
mechanism is Type II.  This conclusion follows from the accretion timescale in a
disk with viscosity coefficient $\alpha$:
\beq
t_{\rm visc}\sim\frac{1}{\alpha}\frac{a^2\Omega}{c_s^2}
\sim4\times10^4\,\,{\rm yr}\,\,\alpha^{-1}_{0.1}\,(M_\star\,a_{100\,{\rm AU}})^{1/2}T_{100\,{\rm K}}^{-1}.
\eeq
One also expects a somewhat larger accreting (active) atmosphere in layered disks with a
dead zone (Gammie 1996).
Similarly, in a self-gravitating marginally stable disk, the characteristic accretion rate is 
\beq
\dot{M}\sim \frac{3 c_s^3}{G}\sim10^{-4}\,\,{\rm M_\odot\,\,yr^{-1}}T_{100\,{\rm K}}^{3/2},   
\eeq
which also contains a strong temperature dependence. 

In addition, for disks embedded in clusters that exceed the ice line temperature
one expects less total mass in planetesimals, and one then expects planetesimal migration 
to be less effective overall.  This implies that the rocky planets that form in 
clusters exceeding the ice line temperature might have different migration history
than their non-cluster analogs for at least three reasons: (1) there might be no Jupiter-mass 
planets to instigate, or regulate, planetesimal migration because the core-accretion
process is quenched; (2) there will be less
mass in planetesimals to begin with because ice cannot condense; 
(3) the viscous time for the disk will be short because $t_{\rm visc}\propto T^{-1}$, which 
is then proportional to the stellar surface density of the cluster as $t_{\rm visc}\propto\Sigma^{-1/4}$
for an optically-thin cluster to  $t_{\rm visc}\propto\Sigma^{-1/2}$ for an optically-thick cluster 
with $T_c>T_{\rm Ice}$.
These countervailing effects ($1+2$ versus 3) need to be considered in assessing the 
demographics of planets found (NGC 6121; see Sections \ref{section:obs} \& \ref{section:discussion})
or not found (47 Tuc) in dense stellar environments, even at fixed metallicity.

\begin{figure*}
\centerline{\includegraphics[width=9cm]{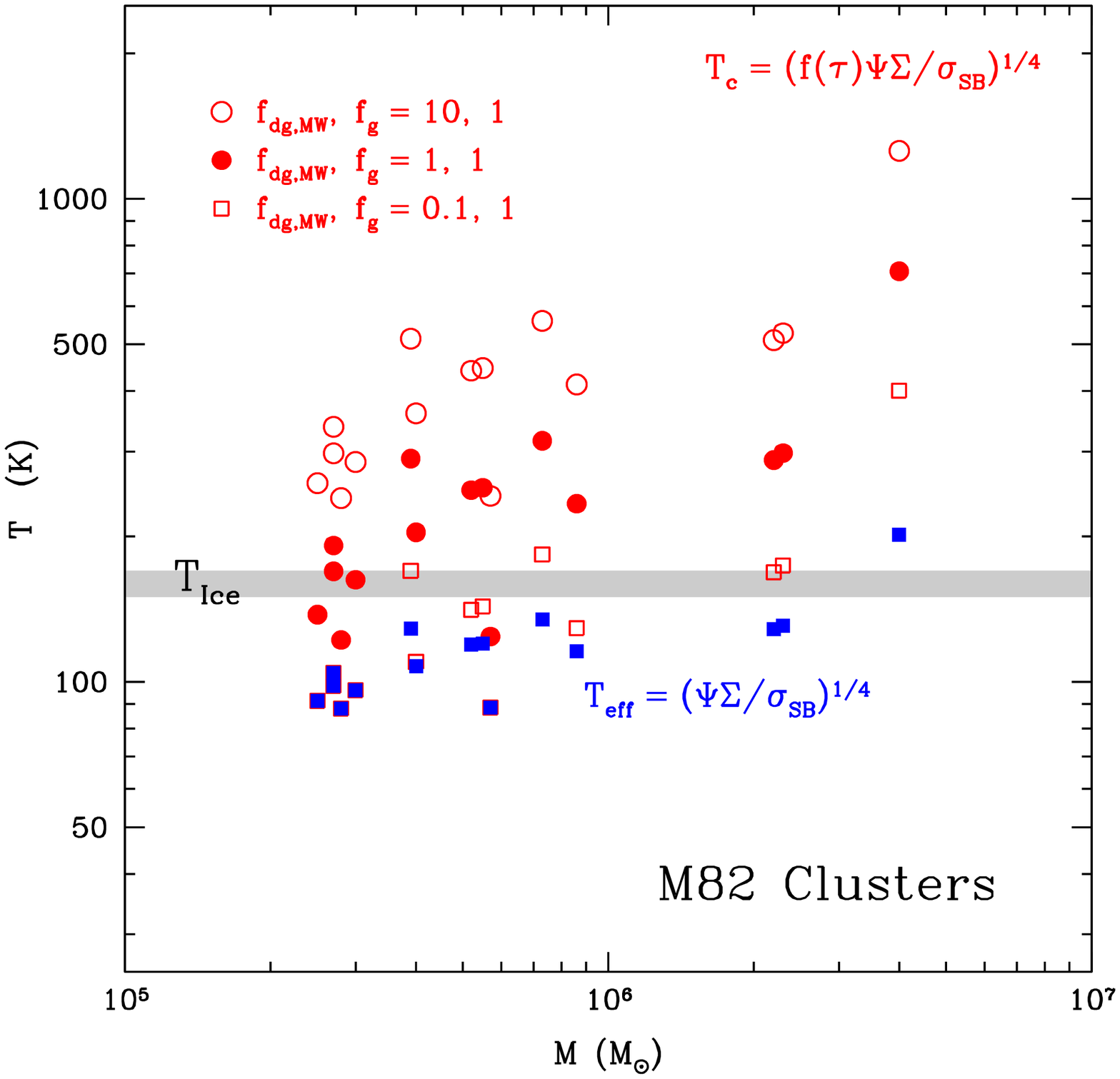}
\includegraphics[width=9cm]{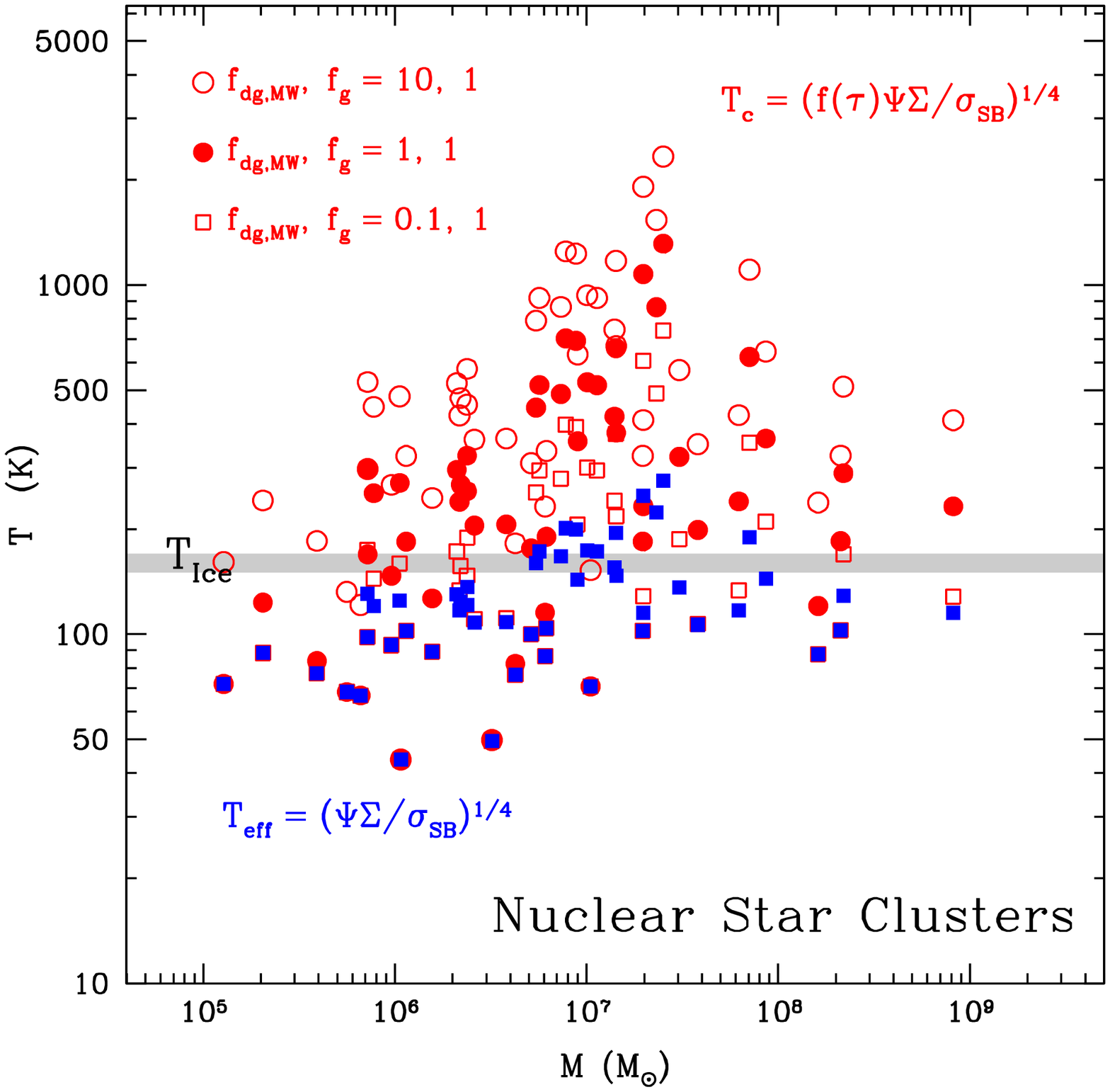}}
\caption{Effective temperature at formation $T_{\rm eff}$ (blue filled squares)
and $T_c$, assuming $\Psi=1500$\,L$_\odot$/M$_\odot$ and 
$f_{\rm dg,\,MW},\,\,f_g=(10,\,\,1)$ (open red circles), $(1,\,\,1)$
(filled red circles), and $(0.1,\,\,1)$ (open red squares), for M82 super star clusters
(left; McCrady \& Graham 2007) and the central star clusters of galaxies  (Leigh et al.~2012).
 Protoplanetary disks embedded in optically-thin clusters reach 
$T_{\rm eff}$ (blue). Disks in optically-thick clusters (red) reach $T_c$ 
during the first $\sim3-4$\,Myr.}
\label{figure:t}
\end{figure*}
\section{Comparison with Data \& Specific Systems}
\label{section:obs}

In this section I use the analytic estimates of the previous 
sections to estimate the temperatures of star clusters, globular 
clusters, and whole galaxies in formation.

\subsection{Super Star Clusters \& Nuclear Star Clusters}

Figure \ref{figure:t} shows $T_{\rm eff}$ (blue filled squares) and $T_c$ (red)
for $f_g=1$ and $f_{\rm dg,\,MW}=10$ (open circles), 1 (filled circles), and 0.1 (open squares),
versus inferred cluster stellar mass $M$, for clusters in M82 (left panel; McCrady \& Graham 2007) 
and for the nuclear star clusters of galaxies (right panel; Leigh et al.~2012).
For the M82 clusters I computed $\Sigma=M/4\pi R^2$, 
using the values of the total mass, velocity dispersion, and crossing time quoted 
in McCrady \& Graham (2007), while for the nuclear clusters I used $M$ and $R$
from Leigh et al.~(2012).  The bolometric flux was then assumed to be 
given by equation (\ref{teff}), assuming $\Psi=1500$\,L$_\odot$/M$_\odot$.
This procedure yields $T_{\rm eff}$ (blue squares) at cluster birth, assuming that the entire stellar mass
formed in less than $\sim3$\,Myr.  The optical depth 
was then estimated using equation (\ref{tau}) for the assumed $f_{\rm dg,\,MW}$, and 
equation (\ref{t}) was solved for $T_c$, giving the red points.

The separation in temperature between $T_{\rm eff}$ and $T_c$ indicates that 
for $f_g=1$ and $f_{\rm dg,\,MW}\gtrsim0.1$, most of the clusters shown in Figure \ref{figure:t} 
are optically-thick on the scale of their half-light radii.
Since $T_c$ measures the minimum temperature in the optically-thick regions of 
protoplanetary disks embedded in these clusters during formation, we see that 
$T_c > T_{\rm Ice}$ in many systems, thus likely suppressing giant planet formation
by core accretion or gravitational instability.

\begin{figure*}[t]
\centerline{
\includegraphics[width=16cm]{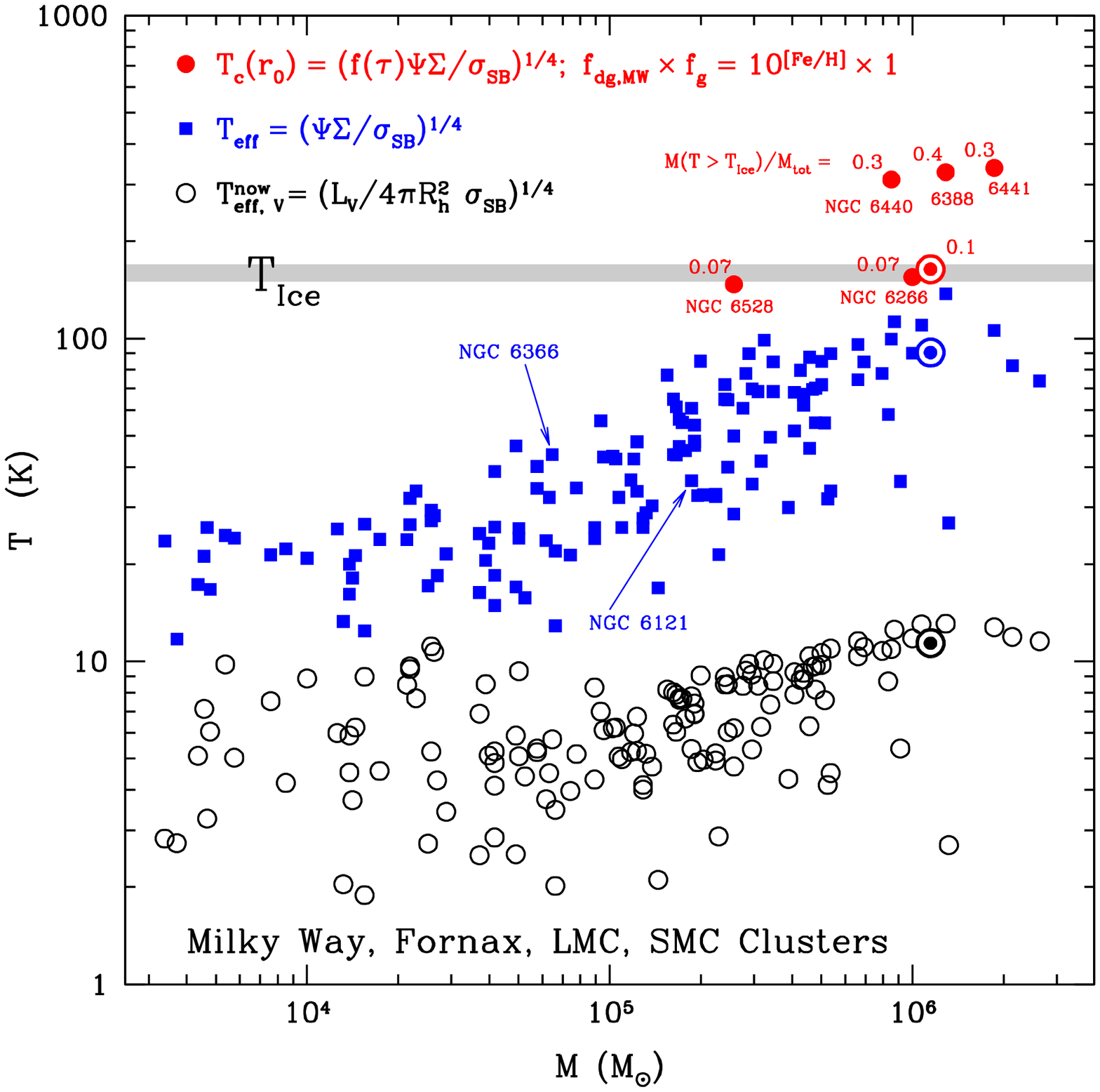}}
\caption{Effective temperature now inferred from $V$-band light
$T_{\rm eff,\,V}^{\rm now}$ (black open circles),
effective temperature at zero-age main sequence $T_{\rm eff}$ (blue filled squares), 
assuming $\Psi=1500$\,L$_\odot/{\rm M}_\odot$,
and {\it central} cluster temperature $T_c$ (see Section \ref{section:globs}) assuming $f_g=1$ and 
$f_{\rm dg,\,MW}=10^{[{\rm Fe/H}]}$ (red) for clusters in
the Milky Way, Fornax, LMC, and SMC (McLaughlin \& Van Der Marel 2005). 
The fraction of the total mass above the ice line temperature in the 
6 optically-thick globular clusters $M(T>T_{\rm Ice})/M$ is noted.
47 Tuc is highlighted (dot+circle). 
 Protoplanetary disks embedded in optically-thin clusters reach 
$T_{\rm eff}$ (blue), whereas disks embedded in optically-thick clusters (red) reach $T_c$ 
during their first $\sim3-4$\,Myr. }
\label{figure:tgc}
\end{figure*}

\subsection{Local Clusters \& Globular Clusters}
\label{section:globs}

In Figure \ref{figure:tgc} I plot temperature versus current stellar mass 
of both young and globular clusters of the Milky Way, Fornax, LMC, and SMC 
(McLaughlin \& van der Marel 2005).
The black open circles show the current $V$-band 
effective temperature computed using $T_{\rm eff,\,now}=(L_V/4\pi R_h^2\sigma_{\rm SB})^{1/4}$,
where $R_h$ is the half-light radius and $L_V$ is the $V$-band luminosity of the cluster 
within $R_h$.  The blue points show the zero-age main sequence  (ZAMS)
effective temperature computed using $T_{\rm eff}=(\Psi\Sigma/\sigma_{\rm SB})^{1/4}$,
where $\Sigma=M/4\pi R_h^2$, $M=\Upsilon_V L_V$ is the total stellar mass within $R_h$, 
$\Upsilon_V$ is the $V$-band mass-to-light ratio, and $\Psi=1500$\,L$_\odot$/M$_\odot$ 
is the assumed ZAMS light-to-mass ratio.  The blue points approximate the effective 
temperature for the cluster in its first $\simeq3$\,Myr and are directly analogous
to the blue squares in Figure \ref{figure:t} for M82 and the nuclear star clusters.  
On timescales longer than $\simeq3$\,Myr, $T_{\rm eff}\propto t^{-1/3}$,
as shown in Figure \ref{figure:l}.

In order to estimate which clusters may have been optically-thick to the 
re-radiated FIR emission from dust, I assumed that 
$f_g=1$ and that the dust-to-gas ratio relative to the 
Galactic value is given by $f_{\rm dg,\,MW}=10^{\rm [Fe/H]}$ (see eqs.~\ref{kappahot} \& \ref{kappacool}).  
Using this prescription for $f_{\rm dg,\,MW}$, I find that only six clusters in the sample are optically
thick on the scale of their half-light radii (all are globulars, labeled in red), and that 
they are only marginally so.  By the same calculation used in the panels of Figure \ref{figure:t}
for $T_c$, only NGC 6440, 6338, and 6441 would have $T_c\sim T_{\rm Ice}$, and although 
47 Tuc, NGC 6266 (M62), and 6526 are optically thick at $R_h$, $T_c<T_{\rm Ice}$ (marginally) by that 
calculation on the scale of $R_h$.

Motivated by the excellent radial profiles presented in McLaughlin \& van der Marel (2005),
and the fairly strong changes in $\Sigma(r)$ and $\rho(r)$ within each cluster,
I was motivated to go a step further and attempt to calculate the fraction of the 
total mass that might reasonably have $T_c>T_{\rm Ice}$.   In order to solve this problem
fully, one would need to solve the radiation transport problem in spherical symmetry, with 
distributed sources, of emissivity $j(r)=\Psi \rho(r)$, subject to the free function of the gas/dust density 
profile, $\rho_g(r)$.  Moreover, one would need to infer the initial stellar density profile 
at ZAMS from the density profile we observe today.
Because these steps are highly uncertain, I made the following crude approximation.
I assume that at every radial position $r$, the total flux carried by that shell 
is the volume integral of the sources $<r$: 
\beq
F(r)=r^{-2}\int_0^r j(r^\prime) r^{\prime\,2} dr^\prime.
\label{fluxr}
\eeq 
At that same position, the total overlying
optical depth is
\beq
\tau(r)=\int_r^\infty \kappa_R\, f_g\,\rho(r^\prime) \,dr^\prime.
\label{taur}
\eeq 
For the required density and emissivity profiles, 
I used the power-law ($\gamma$-) models presented by McLaughlin \& van der Marel (2005)
for each cluster, which yield the three-dimensional emissivity profile 
\beq
j(r)=\Psi\rho_0\,\left[1+(r/r_0)^2\right]^{-\gamma/2},
\eeq
where $\rho_0$ and $\gamma$ are provided for each cluster by the authors.
I then assumed $f_g=1$ and $f_{\rm dg,\,MW}=10^{\rm [Fe/H]}$, and 
solved equation (\ref{t}) for $T_c(r)$ using $F(r)=\sigma_{\rm SB}T_{\rm eff}$ and $\tau(r)$
in equations (\ref{fluxr}) and (\ref{taur}).
This procedure accounts for the fact that both the 
total flux and optical depth vary strongly with $r$, and allows one to make an estimate of the central temperature
of the cluster where $F(r)$ may be very different from $F(R_h)$, and $\tau(r)$ may be much larger than  $\tau(R_h)$.

The solid red dots in Figure \ref{figure:tgc} show the resulting {\it central} cluster temperature.  The 
small red number shows the fraction of the mass that has temperature larger than 
$T_{\rm Ice}\simeq150\,$K, $M(T>T_{\rm Ice})/M$.   This calculation thus implies that only 
$\sim0.1-0.4$ of the total mass of these clusters exceeds the ice-line temperature.  Generically,
it is the central regions, where the flux is still very high, and the overlying optical depth is 
very large, where $T_c(r)>T_{\rm Ice}$ obtains.
I emphasize that this procedure is very crude since it (1) is subject to the uncertainties of a free function,
$\rho_g(r)$, which I have specified to be $f_g \rho(r)$, (2) does not in fact solve the 
transport equation with the non-gray opacity of dust, and (3) assumes that $\rho(r)$ is unchanged
from its initial distribution.
Nevertheless, this procedure allows one to explore the obvious change in $F(r)$ and $\tau(r)$,
which undoubtedly lead to temperature profiles in clusters at formation.

I discuss a number of specific systems further in Section \ref{section:discussion},
but here note that none of the clusters highlighted in red, with $M(T>T_{\rm Ice})/M>0$
are thought to have undergone core collapse since birth.

\subsection{Galaxies \& AGN}
\label{section:galaxies}

The formalism discussed so far for star clusters can be generalized to 
continuously star-forming galaxies by taking $F=\varepsilon\dot{\Sigma}_\star c^2$, 
where $\dot{\Sigma}_\star$ is the star formation rate per unit area, and $\varepsilon$
is an IMF-dependent constant of order $5\times10^{-4}-10^{-3}$ for standard IMFs.
The effective and central temperatures of galaxies in formation are then
(compare with eqs.~\ref{teff} and \ref{tchot}; see the related discussion in 
Thompson, Quataert, \& Murray 2005)
\beq
T_{\rm eff}\simeq102\,\,{\rm K}\left(\varepsilon_{\,-3}\,\dot{\Sigma}_{\star,\,3}\right)^{1/4}
\label{teffgal}
\eeq
and 
\beq
T\sim170\,{\rm K}\,\left[f_{\rm dg,\,MW}\,\varepsilon_{-3}\,\Sigma_{g,\,4}
\dot{\Sigma}_{\star,\,3}\right]^{1/4},
\label{tgal}
\eeq
respectively, where $\varepsilon_{\,-3}=\varepsilon/10^{-3}$, 
$\dot{\Sigma}_{\star,\,3}=\dot{\Sigma}_{\star}/10^3$\,M$_\odot$ yr$^{-1}$ kpc$^{-2}$,
and $\Sigma_{g,\,4}=\Sigma_g/10^4$\,M$_\odot$ pc$^{-2}$.
Equation (\ref{tgal}) assumes $\tau\gg1$ and equation (\ref{kappahot}).\footnote{Using
$\varepsilon=5\times10^{-4}$ decreases eqs.~\ref{teffgal} and \ref{tgal} to 
86\,K and 140\,K, respectively.}

\begin{figure*}
\centerline{\includegraphics[width=12cm]{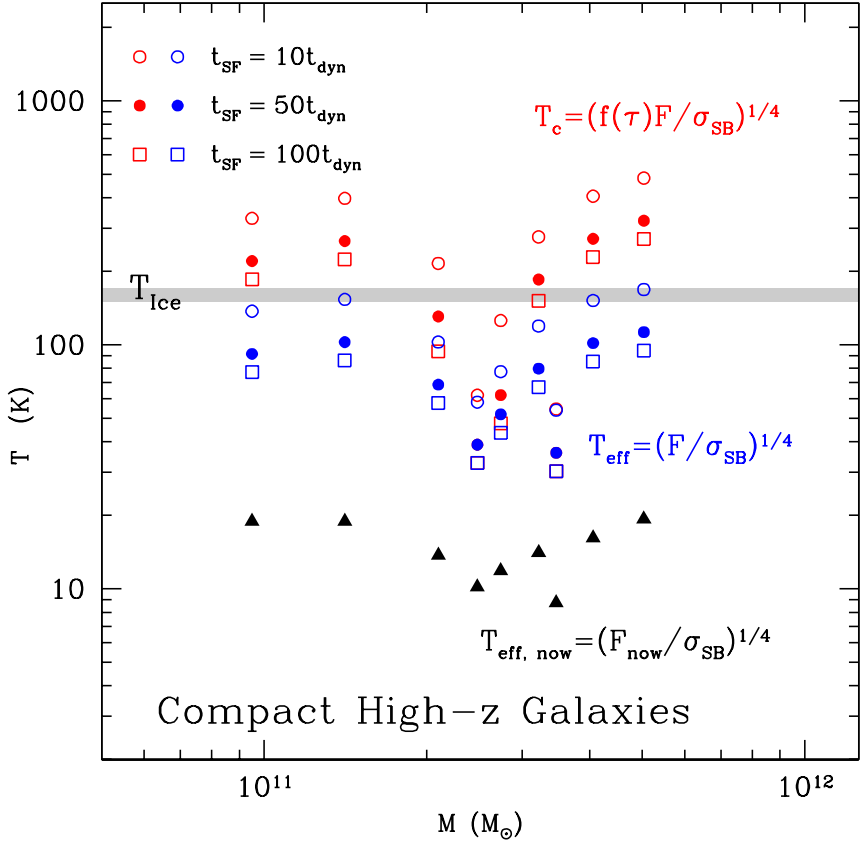}}
\caption{Similar to Figure \ref{figure:t}, but for 
compact high-$z$ galaxies (Van Dokkum et al.~2008; Kriek et al.~2008),
assuming that $f_g=f_{\rm dg\,MW}=1$, and that 
their stars formed in single burst of duration $10\times$ (open circles), $50\times$ (filled circles), 
and $100\times t_{\rm dyn}$ (open squares) (see Section \ref{section:galaxies}).  Blue and red
show $T_{\rm eff}$ and $T_c$, respectively.  $T_{\rm eff,\,now}$ is shown as
the filled black triangles.}
\label{figure:van}
\end{figure*}

As an application, in Figure \ref{figure:van} I show temperature as a function of stellar mass 
estimated for the compact high-$z$ galaxies 
from the sample of Van Dokkum et al.~(2008) and  Kriek et al.~(2008).
The filled black triangles show the current effective temperature 
$T_{\rm eff,\,now}=(M/\Upsilon 4\pi R_{\rm eff}^2\sigma_{\rm SB})^{1/4}$,
where $\Upsilon$ is the current mass-to-light ratio
of the stellar population found by  Kriek et al.~(2008), and $M$ and $R_{\rm eff}$ are
the total stellar mass and effective radius.  I then compute the dynamical timescale
$t_{\rm dyn}\sim(G\rho)^{-1/2}$  for each galaxy and assume that the star formation timescale is 
$t_{\rm SF}=10\times$ (open squares), $50\times$ (filled circles), $100\times t_{\rm dyn}$ (open circles).  
The flux at formation is then $F=\varepsilon \Sigma c^2/t_{\rm SF}$, where I approximate 
$\Sigma=M/4\pi R_{\rm eff}^2$.  The blue points
show $T_{\rm eff}=(F/\sigma_{\rm SB})^{1/4}$, the effective temperature during formation,
while the red points show the solution to equation (\ref{t}) assuming that $\Sigma_g=\Sigma$ ($f_g=1$) and
$f_{\rm dg,\,MW}=1$, which approximates the estimate of equation (\ref{tgal}) at high optical depths.

Although highly simplified, these estimates imply that all of these systems were
$\gtrsim 30$\,K at formation and that many were $\gtrsim 70$\,K.  And
$\sim6$ of the systems shown here likely 
attained temperatures in excess of $T_{\rm Ice}$, almost certainly suppressing
planet formation by core accretion or gravitational instability.

Similar estimates can be made for actively star forming galaxies.  Equations (\ref{teffgal})
and (\ref{tgal}) show that it is the most gas-rich and highly star forming systems that 
are likely to suppress planet formation.  These include the inner regions of 
Arp 220 and other local ULIRGs, high-$z$ submillimeter selected sources, and the 
highly star-forming clumps within high-$z$ galaxies like that in Q2346-BX 482
(Genzel et al.~2008; Murray et al.~2010). 
One can use the Schmidt law as derived by
Kennicutt (1998) to connect $\dot{\Sigma}_{\star}$ and $\Sigma_g$ in actively star-forming
galaxies.  Equation (\ref{tgal}) then allows an estimate of the critical gas surface density above
which $T$ exceeds $T_{\rm Ice}$:
\beq
\Sigma_{g,\,{\rm Ice}}\sim2\times10^4\,\,{\rm M_\odot\,\,pc^{-2}}\,\,T_{\rm Ice,\,150\,K}^{1.7}\,\,\varepsilon_{\,-3}^{-0.4}.
\label{sigmaicegal}
\eeq
Such surface densities are in fact achieved in the inner regions of Arp 220
(Downes \& Solomon 1998).    Matsushita et al.~(2009) show that this starburst is optically-thick
at 435\,$\mu$m, thus justifying the estimate that $\tau>1$, and Downes \& Eckart (2007) find a true dust 
temperature of $170$\,K for the inner disk.  Herschel observations presented by Rangwala et al.~(2011)
show that the dust has optical depth of $\sim5$ at 100\,$\mu$m, and multiple molecular 
line diagnostics indicate the presence and pervasiveness of hot dust and gas (see also Ott et al.~2011).

Again, equation (\ref{teffgal}) sets the fundamental temperature scale for the 
protoplanetary disks embedded in highly star-forming galaxies and starbursts.
Since star formation surface density and gas/dust surface density are highly 
correlated through the Schmidt law, regions with $\dot{\Sigma}_{\star}\gtrsim10^3$\,M$_\odot$ yr$^{-1}$ kpc$^{-2}$
will have $\tau>1$ and the temperature of the system will be higher than the 
effective temperature by a factor of $\tau^{1/4}$.  One sees from these simple estimates
that, although extreme, many systems may reach this limit.  

Since the compact high-$z$
galaxies presented in Figure \ref{figure:van} may be the cores of present-day ellipticals
(Hopkins et al.~2009), our expectation for the prevalence of planets in these metal-rich 
systems should be adjusted accordingly strongly downward.  Although rare in the local universe, 
the conditions producing $\Sigma_{\rm Ice}$ in equation (\ref{sigmaicegal}) were more 
common in the high-$z$ universe, and the subsequent high temperatures and suppression of 
planet formation via gravitational instability or core accretion should affect the 
present day planetary mass density. 

Finally, AGN activity can also heat the medium above $T_{\rm Ice}$.
For a BH luminosity of $L=L_{46}10^{46}$\,ergs s$^{-1}$,
$T_{\rm eff}\gtrsim T_{\rm Ice}$ for $R\lesssim50\,{\rm pc}\,\,L_{46}^{1/2}T_{\rm Ice,\,150\,K}^{-2}$.
Assuming an isothermal sphere for the stellar 
mass distribution in the central regions of galaxies, this corresponds to 
$M(<R)\sim\sigma^2 R/G\simeq5\times10^8\,\,{\rm M_\odot}\,\,(\sigma/200\,{\rm km/s})^2 (R/50\,{\rm pc})$,
where $\sigma$ is the velocity dispersion, about $1$\% of the total bulge mass.  Supermassive BHs 
may undergo significant growth during rapid star formation episodes on multi-kpc scales.
The planet population may thus be affected by both the episode of BH growth and the star formation
itself.

\section{Discussion}
\label{section:discussion}

\subsection{Caveats, Complications, \& Uncertainties}
\label{section:uncertain}

{\it Radiation Transport ---} Equation (\ref{t}) is highly approximate. It applies
only in the limit of a an overlying column of matter without 
sources, makes the Eddington approximation, and assumes both an infinite plane-parallel atmosphere
and gray opacity.   A real embedded star cluster has a distribution
of sources (stars), whose light is reprocessed locally into
the infrared by absorption, and which contributes to the 
local flux in the atmosphere. The radial distribution of sources
$j(r)=\Psi \rho(r)$, the distribution of dusty 
gas with respect to those sources, $f_g(r)=\Sigma_g(r)/\Sigma(r)$,
 the variation in the IMF as a function of radius 
$\Psi(r)$, the variation in the dust-to-gas ratio $f_{\rm dg,\,MW}(r)$
and thus the opacity, will all affect the radial run of temperature with optical depth,
$T(\tau)$.    

A detailed exploration of these effects is beyond the scope of this work, but I 
note here that the formalism of Hubeny (1990) for distributed sources 
in plane-parallel atmospheres can
be generalized to spherical systems with non-gray opacities (see, e.g., Whelan et al.~2011).
However, more detailed radiation transport models should be coupled to 
a self-consistent calculation of the dynamics of star-forming and disrupting clusters since 
$f_g(r)=\Sigma_g(r)/\Sigma(r)$ would otherwise be a free function.  In addition,
since star formation may be spatially correlated within individual star clusters,
the radiation transport problem is likely to be further complicated by multi-dimensional
effects (e.g., Krumholz et al.~2012), and, further, by the dynamics of merging sub-clumps.

{\it The Optical Depth \& Turbulence ---} Recent calculations suggest that massive clusters may
be disrupted by radiation pressure on dust grains, and this feedback mechanism has been suggested
as a way to regulate star formation on galaxy scales
(Murray et al.~2005; Thompson et al.~2005; Krumholz \& Matzner 2010;
Murray et al.~2010; Andrews \& Thompson 2011; 
Hopkins et al.~2012).  The parameter space over which this feedback mechanism
may dominate largely overlaps with the parameter space discussed in this work (e.g., 
$\Sigma \gtrsim 10^3$\,M$_\odot$ pc$^{-2}$ or $\dot{\Sigma}_\star\gtrsim10^3$\,M$_\odot$ yr$^{-1}$ kpc$^{-2}$).
If the gaseous medium is highly turbulent,
and in particular, if the radiation field itself drives that turbulence, then the 
effective optical depth for the reprocessed IR radiation may be less than the
simple estimate of equation (\ref{tau}) (see Murray et al.~2005; Hopkins et al.~2012).
Recent simulations by Krumholz \& Thompson (2012) of sustained radiation pressure 
driven Rayleigh-Taylor convection (in 2D) in fact suggest that in these environments the value of the 
effective optical depth is decreased by a factor of $\sim2-5$, as low optical depth channels through the 
overlying medium are formed and the radiation field more efficiently escapes.
In principle, this should decrease the temperature below the approximation of 
equation (\ref{t}), but this effect has yet to be quantified in the super star cluster
environment.

{\it IMF ---}  
Observations imply that the light-to-mass ratio of a ZAMS stellar population may 
vary. The value of $\Psi$ I scale to throughout this work, $1500$\,L$_\odot$/M$_\odot$,
is approximately appropriate for a Salpeter IMF from $1-100$\,M$_\odot$.  More precisely, 
$\Psi(t=0)\simeq1400$ and $\Psi(t=3{\rm\,Myr})\simeq1900$\,L$_\odot$/M$_\odot$  (see Fig.~\ref{teff}).  
For limits of $0.1-100$\,M$_\odot$, unrealistic for 
the low-mass stellar field population (e.g., Kroupa et al.~1993; Kroupa 2001; Chabrier 2003),
$\Psi(t=3{\rm\,Myr})$ decreases to $\simeq750$.  For a Kroupa IMF 
$\Psi(t=3{\rm\,Myr})\simeq1200$\,L$_\odot$/M$_\odot$.  A somewhat more bottom-heavy IMF below
1\,M$_\odot$,
but not as steep as Salpeter, as observed in the Galactic bulge by Zoccali et al.~(2000), 
would yield similar values for $\Psi$.
Stolte et al.~(2006) measure the present day mass function in the young 
compact Galactic star cluster NGC 3603 and find a slope of $N\propto M^{-0.9\pm0.15}$
from $\sim1-20$\,M$_\odot$.  In Arches, Stole et al.~(2005) find a similar slope 
at high masses, but with a significant turnover at $\sim6-7$\,M$_\odot$.
Such a flatter than Salpeter slope at high masses, if extended to $\sim100$\,M$_\odot$,
and taking Kroupa from $0.1-1$\,M$_\odot$, 
would increase $\Psi$ over our nominal value by a factor of $\sim1.5-2$.

The value of $\Psi$ and its time dependence also change if the stellar population is rotating
(Ekstr{\"o}m et al.~2012).  Levesque et al.~(2012) (e.g., their Fig.~6) show that 
$\Psi$ reaches a maximum at $\simeq3.5$\,Myr, effectively extending the life of the 
massive stars that dominate the bolometric output of the stellar population by $\sim0.5-1$\,Myr
over the non-rotating population.
At an age of 5\,Myr, the rotating population is $\simeq1.6$ times brighter for the 
same stellar mass, and at 10\,Myr it is $\simeq1.3$ times brighter.  Rotation would 
thus increase the temperatures of the protoplanetary disks of star clusters, all else
being equal.

The normalization, shape, and high-mass slope of the IMF may also change as a function 
of cluster mass and surface density.  Murray (2010) presents an argument that it is
precisely the FIR optically-thick star clusters with the highest values of $\Sigma$ 
and $\Sigma_g$ that will deviate most strongly from a standard IMF, with larger $\Psi$.
Such a correlation --- $\Psi \propto\Sigma$ or $M$ to some power --- would strongly increase 
$T_{\rm eff}$ and $T_c$ with respect to equations (\ref{teff}) and 
(\ref{tchot}) in the highest surface density systems.  As an 
example,  super star cluster M82-F was argued by Smith 
\& Gallagher (2001) to be strongly inconsistent with a standard Kroupa IMF, and 
requires a low-mass cutoff of $\sim2-3$\,M$_\odot$.  Such an IMF would  
increase $\Psi$ for the ZAMS population, increasing the inferred temperature
in the first few Myr of evolution.  A strongly bottom-heavy IMF, as has been claimed 
in elliptical galaxies (Van Dokkum \& Conroy 2010, 2011; Wegner et al.~2012), 
and which may be a function of velocity dispersion (Spiniello et al.~2012), 
would go the other way, decreasing $\Psi$ by a factor of $\sim2$ or more.

Finally, there is evidence for a bottom-heavy IMF in elliptical galaxies
(Van Dokkum \& Conroy 2010, 2011; Wegner et al.~2012), which 
may be a function of velocity dispersion (Spiniello et al.~2012), and which
would decrease the temperature by decreasing $\Psi$ relative to the fiducial
value used throughout this work.

{\it Time Dependence ---} The central temperature of the star cluster is likely to 
change as a function of time, as the gas is used up, the stellar population grows,
the stellar population evolves dynamically, and as the cluster is disrupted, ejecting 
the remaining gas and perhaps unbinding or expanding the stellar distribution.
A more realistic treatment would couple the evolution of $T$ throughout the 
cluster with a dynamical calculation of cluster disruption, as in the simplified 
models of Murray et al.~(2010), or in the (much lower mass) cases of numerical 
star cluster formation with realistic thermodynamics 
(e.g., Bate 2010, 2011, 2012; Krumholz et al.~2012).

In any case, since the timescale for the embedded cluster phase is likely to 
be a multiple of the dynamical timescale, the average inner cluster temperature 
may decrease rapidly from $T_c$ to $T_{\rm eff}$ as the dusty column trapping 
the radiation disperses.  Since this timescale is of order the timescale 
required for planet formation $\sim$\,Myr, planet formation may be fully 
quenched by the fact that the medium is above $T_{\rm Ice}$, or merely delayed
for $\sim$\,Myr until cooler climes prevail, $T_c < T_{\rm Ice}$ and giant planets
form.  

A related point is that the planet formation timescale is
shortened in the cluster environment because of the 
increased heating of the surface layers by the cluster irradiation (Section \ref{section:accretion}),
radioactivity of the (potentially) large-scale 
starburst environment (Lacki 2012a), the very large cosmic ray ionization rates
and gamma-ray fluxes expected in those environments (Lacki et al.~2010; Papadopolous 2010;
Papadopolous et al.~2011; Lacki 2012b), or 
the UV irradiation from nearby massive stars (see Section \ref{section:introduction}).  
All act to increase the 
size of the active accreting zone in layered disks (Gammie 2001), and hence
speed both the accretion of the disk onto the host star, 
and potentially the migration of planets in formation.

\subsection{RV \& Transit Surveys in Galactic Globular Clusters \& Open Clusters}
\label{section:rvcluster}

As discussed in Section \ref{section:uncertain} there are substantial uncertainties in 
the estimates of $T_{\rm eff}$ (blue), and especially $T_c$ (red) in Figures
\ref{figure:t}, \ref{figure:tgc}, and \ref{figure:van}.  Even so, they
 make it clear that globular clusters, star clusters
in the local universe, compact high-$z$ galaxies, and nuclear star clusters  
 were born very hot, and that many of 
these systems were likely optically thick at birth, leading to $T_c$ above the ice line 
temperature $T_{\rm Ice}$.   
Given the discussion of Section \ref{section:formation}
one would thus expect that these systems should not have been
able to form gas- and ice-giants by core-accretion during their first $\sim1-10$\,Myr
of evolution, that formation of similar systems by gravitational 
instability would have been prohibited in all but the most massive disks, 
and that there should not have been a delivery mechanism for water to 
terrestrial planets on these short timescales.

In Figure \ref{figure:tgc}, I highlight 47 Tucanae
(${\rm [Fe/H]=-0.76}$; McLaughlin \& van der Marel 2005),
which has been the subject of an intensive observational campaign to find
giant planets via transits (Gilliland et al.~2000; Weldrake et al.~2005). 
Using our nominal numbers for 47 Tuc --- $f_{\rm dg,\,MW}=10^{[{\rm Fe/H}]}$ and $f_g=1$ ---
I find that $\sim0.1$ of the mass had $T_c>T_{\rm Ice}$.  Changes of a
factor of $2$ in $f_g$, $\Psi$, or the effective radius for the stars
change this fraction from $\sim0.1$ to $\sim0-1$.

Given these substantial uncertainties, it is worth considering the implications
of the hypothesis that the lack of planets observed in 47 Tuc is a result of 
a high temperature at birth.
There are several possibilities, depending on whether or not 
planets form predominantly by core accretion or gravitational instability,
and what the true frequency of giant planets is as a function of metallicity,
independent of birth environment.  If planets form predominantly
by core accretion, and if the lack of 
giant planets in 47 Tuc in the survey of Gilliland et al.~(2000) is indeed 
entirely because $T_c>T_{\rm Ice}$ in its first Myrs of evolution,
then I predict that  globular clusters with similar metallicity, but lower $\Sigma$  
will harbor giant planets if they had $T_c < T_{\rm Ice}$. For this reason, 
I highlight NGC 6366 in Figure \ref{figure:tgc}, 
which has a metallicity comparable to 47 Tuc
(${\rm [Fe/H]=-0.82}$; McLaughlin \& van der Marel 2005), but an average 
stellar surface density that is $\simeq10$ times smaller.   
NGC 6366 is the only 
relatively high-metallicity globular cluster with $\Sigma$ substantially
below 47 Tuc cataloged in McLaughlin \& van der Marel (2005), and it is 
thus an ideal case for testing planet formation 
as a function of surface density at nearly fixed metallicity.

A similarly important test is the globular cluster NGC 6121 (M4), where a  
2.5\,M$_{\rm Jupiter}$ mass planet has been inferred 
by Sigurdsson et al.~(2003).  NGC 6121 has ${\rm [Fe/H]=-1.2}$, 
and a simple extrapolation of Fischer \& Valenti's correlation between 
planet frequency and metallicity would suggest that it should harbor 
very few giant planets.  Is this planet's existence in conflict
with our hypothesis for 47 Tuc?  No.  As shown in Figure \ref{figure:tgc},  
the effective temperature of NGC 6121 was likely 
substantially below $T_{\rm Ice}$ at birth as a result of its low metallicity and
surface density with respect to 47 Tuc.  

The planet in NGC 6121 can also be thought of as a test for gravitational
instability, which would then have to operate in a radiation environment
of $T_{\rm eff}\approx40$\,K, implying a massive protoplanetary disk (see Section \ref{section:grav}).

Several additional systems are worth discussing in this context.  In particular,
the more metal-rich globular clusters of the Galaxy listed in McLaughlin \& van der Marel (2005)
are NGC 6440, 6441, and 6388 with ${\rm [Fe/H]=-0.34}$, $-0.53$, $-0.60$,
respectively.  These are the hottest three systems  $T_c$ significantly above $T_{\rm Ice}$, 
shown in red in Figure \ref{figure:tgc}.  Therefore, these systems also present
a test of the core-accretion scenario, and this paper.  If 47 Tuc indeed formed 
no (or few) gas- and ice-giants because $T_c > T_{\rm Ice}$, then NGC 6440, 6441, and 6388
should be similarly devoid.  

I emphasize that such a set of tests would be complicated
by the many effects that may inhibit planet formation in dense clusters (see Section \ref{section:introduction}).
Nevertheless, comparing the planet population at different metallicity {\it and}
at different stellar surface density is one of the few ways of understanding these
joint dependencies.

Finally, note that  one of the most compact and metal-rich of the Galactic 
open clusters is NGC 6791, which has been 
the subject of a number of transit surveys (Section \ref{section:introduction}),
has $T_{\rm eff}({\rm ZAMS})\simeq20$\,K and $\tau_R\simeq0.04$, using $M\sim5000$\,M$_\odot$, 
$R\sim5$\,pc, $f_{\rm dg,\,MW}=10^{[{\rm Fe/H}]}$, ${\rm [Fe/H]}=0.3$, and  taking $f_g=1$
(Platais et al.~2011 and references therein), and thus its planet population 
should not be strongly affected by the effects described in this work.

\subsection{Observations of Protoplanetary Disks in Dense Stellar Environments}

The simple calculations shown in Figure \ref{figure:cg} and the discussion of 
Section \ref{section:embed} indicate that the vertical structure of protoplanetary 
disks in dense stellar environments should be strongly modified from that derived for
passive or active disks without external irradiation.  The work of 
Stolte et al.~(2004), (2006), and (2010) and others use NIR excess to investigate the properties and incidence of 
disks within clusters like NGC 3603 and Arches.  More detailed calculations of disks
with self-consistent thermodynamics and structure are warranted for observations
now and coming (e.g., JWST) of stars and their disks in the most massive and 
dense Galactic clusters.

As an example, using the data collected in Table 2 from Portegies-Zwart et al.~(2010),
Arches has $T_{\rm eff}\simeq100$\,K and $T_c\simeq180$\,K for $f_g=f_{\rm dg,\,MW}=1$.
NGC 3603 has lower temperature, and is not optically-thick for $f_g=f_{\rm dg,\,MW}=1$,
but still has $T_{\rm eff}\simeq70$\,K.  Trumpler 14 and Westerland 1 have similar stellar 
surface densities and thus similar temperatures to NGC 3603. Figure \ref{figure:cg} and the scalings of Section
\ref{section:embed} imply that the protoplanetary disks in these systems should be 
strongly modified by their radiation environment.

\subsection{The Maximum Temperature of Dense Stellar Systems}
\label{section:extreme}

The maximum stellar surface density observed in star clusters, elliptical
galaxies, globular clusters, and other dense stellar systems is 
(Hopkins et al.~2010)
\beq
\Sigma_{\rm max}\sim10^{5}\,\,{\rm M_\odot\,\,pc^{-2}}.
\label{sigmamax}
\eeq
If this stellar surface density was formed in a single burst,
$F_{\rm max} =1.5\times10^{14}\,\Psi_{1500}\,{\rm L_\odot\,\,kpc^{-2}}$,
$T_{\rm eff}^{\rm max}\sim180\,{\rm K}\,\,\Psi_{1500}^{1/4}$,
and 
$\tau^{\rm max}_R\sim100\,f_{\rm dg,\,MW}\,f_g$.  Taking the 
maximum dust-to-gas ratio observed in galaxies to be $f_{\rm dg,\,MW}\sim10$
(Mu{\~n}oz-Mateos et al.~2009)
implies that the maximum temperature it might be possible to achieve
within a star cluster is 
\beq
T_{\rm max}\sim10^3\,{\rm K}\,\,(f_{\rm dg,\,MW}\,f_g\,\Psi_{1500}/10)^{1/4}.
\eeq
Compare with Figure \ref{figure:temp}.
This temperature is equivalent to the effective insolation
at a distance of $\sim0.5$\,AU around star of Solar luminosity,
and it is perhaps the largest value of the IR cluster
insolation that can be obtained in the central regions of dense 
stellar systems.   Although $T_{\rm max}$ exceeds the condensation
temperatures of magnetite ($\simeq370$\,K), troilite ($\simeq700$\,K),
and approaches the condensations temperatures of feldspar and forsterite 
this value is still substantially
below the sublimation temperatures of the primary refractory condensates, 
and so it should not be possible to directly inhibit the formation of 
rocky planets.

\subsection{The Fraction of All Star Formation
Occurring above a Given Stellar Surface Density}
\label{section:universe}

If massive dense star clusters and galaxies in formation inhibit the formation
of gas and ice giant planets by core-accretion, as implied by Sections \ref{section:formation}
and \ref{section:obs}, then it is worth asking about the fraction of all star formation
that has occurred above the critical surface density $\Sigma_{\rm Ice}$, such that 
$T_c>T_{\rm Ice}$, as estimated in equation (\ref{sigmaice}).  Such a calculation
requires several ingredients.  Equation (\ref{tchot}) implies that 
\beq
d\ln T_c=\frac{1}{2}d\ln\Sigma+\frac{1}{4}d\ln Z+\frac{1}{4}d\ln \Psi,
\eeq
where I have assumed that $f_g\sim{\rm const}$, 
and $Z$ is the metallicity normalized such that $f_{\rm dg,\,MW}\propto Z$,
which enters into the dust opacity per unit gram of gas in equation (\ref{kappahot}).
Over the history of most of the star formation in the universe, one expects 
$Z$ to vary by $\sim2-3$\,dex.  The total range in $\Psi$ is 
likely smaller, only varying by (at most) a factor of $\sim3-5$, and it may be 
approximately constant.  The total range in $\Sigma$ is quite large, spanning 
from $\sim10^2-10^{5}$\,M$_\odot$ pc$^{-2}$.
Naively, then, the fraction of all stars formed above $\Sigma_{\rm Ice}$ depends 
primarily on
the cluster mass function $dN/dM$, but also $dN/dR$ and $dN/dZ$, with a much smaller dependence on 
$dN/d\Psi$. 

Typical functions for $dN/dM$ for clusters yield $dN/dM\propto M^{-2}$, but with a 
cutoff at high masses that is dependent on the galaxy observed.  It is natural 
to take the maximum mass to be a fraction of the Toomre mass in a marginally stable 
galactic disk of gas surface density $\Sigma_g$ and scale-height $h$, $\pi h^2 \Sigma_g$,  
but the star formation efficiency as a function of the collapsed gas cloud may 
vary as a function of its mass. Similarly, the relationship between cluster mass
or richness and its size is likely not a simple power-law over the range 
from $M\sim10^2-10^7$\,M$_\odot$ (e.g., see discussions in Murray 2009; Murray et al.~2010; Adams 2010).
In particular, $R\propto M^{0.3-0.4}$ for $M<10^4$\,M$_\odot$ (Lada \& Lada 2003) and 
$R\propto M^{0}$ for $10^4<M<10^6$\,M$_\odot$ with $R\sim2$\,pc.  For larger mass clusters 
the relationship apparently steepens, with $R\propto M^{0.6}$ ($M>10^6$\,M$_\odot$)
(Walcher et al.~2005; Murray 2009).

Given these various uncertainties and unknowns I take a simple tack and make an
estimate for an average galaxy that dominates the star formation rate of the 
universe at redshift $\sim1-2$.  Such star-forming galaxies have gas surface densities $\sim10-100$
times the Galaxy, with commensurately higher star formation rates per unit area. 
Their metallicities 
are likely sub-Solar on average by a factor of a few.  If one assumes that a typical 
cluster size is  that $dN/dM\propto M^{-2}$, then the fraction 
of all star formation occuring with $T_c>T_{\rm Ice}\simeq150\,$K is 
just $f(T_c>T_{\rm Ice})=\ln(M_{\rm max}/M_{\rm crit})/\ln(M_{\rm max}/M_{\rm min})$
where 
\beq
M_{\rm crit}\simeq2.5\times10^5\,\,{\rm M_\odot}(R/2\,{\rm pc})^2(f_{\rm dg,\,MW}\,f_g\,\Psi_{1500})^{-1/2}
\eeq
is the critical cluster mass above which $T_c>T_{\rm Ice}$ (see eq.~\ref{sigmaice}),
and $M_{\rm min}$ and $M_{\rm max}$ are the minimum and maximum cluster masses, respectively.
Taking $R=1$, 2, and 4\,pc, $f_{\rm dg,\,MW}=f_g=\Psi_{1500}=1$,
$M_{\rm min}=10$\,M$_\odot$, and $M_{\rm max}=10^7$\,M$_\odot$, 
 $f\sim37$\%, 27\%, and 17\% of stars are born in 
clusters with $T_c>T_{\rm Ice}$, respectively.
For $f_{\rm dg,\,MW}=0.1$, these fractions decrease to $\sim28$\%, 18\%, and 8\%, 
respectively.  

In this context, it is worth noting the recent work of Gieles et al.~(2010) who
provide an explanation for the $M-R$ relation for stellar systems above $\sim10^4$\,M$_\odot$
(see their Fig.~3).  They show that dynamical processes increase the half-light radii ($R$)
of clusters of Gyr timescales.  Their initial configuration starts with 
$M\sim 10^6\,{\rm M_\odot}\,\,(R/{\rm pc})^{2}$, a constant surface density 
at all masses equivalent to $\Sigma\simeq8\times10^4$\,M$_\odot$ pc$^{-2}$ and
an effective temperature at zero-age main sequence of $T_{\rm eff}\simeq170$\,K
from equation (\ref{teff}). Since the timescale for significant dynamical evolution in 
these simulations is 10\,Myr or longer I expect systems that start with 
this value for $\Sigma$ to have $T_c>T_{\rm Ice}$ throughout the crucial 
planet formation epoch.  If such an initial function for the  $M-R$ relation
was in fact generically valid for $M>10^4$\,M$_\odot$, the arguments presented
in this paper would imply that more than 50\% of all stars were born in 
an environment hostile to formation of gas and ice giant planets via core accretion 
or gravitational instability.

Finally, in a study of the structure of elliptical galaxies, and compact galaxies 
at high redshift, Hopkins et al.~(2009) estimates that 
the total mass in present-day ellipticals with stellar surface 
densities in excess of $\Sigma_{\rm Ice}$ is $\sim10-20$\%.

\section{Conclusions}
\label{section:conclusion}

The basic conclusions of this work are as follows:

1.~Massive star clusters 
with standard IMFs that form in a single burst of star formation 
are hot for a time comparable to the planet formation timescale,
reaching fluxes equivalent to an effective temperature of $\sim50-150$\,K
for $\sim3-4$\,Myr (Figure \ref{figure:l}; eq.~\ref{teff}).

2.~Embedded dusty star clusters reach a temperature larger than $T_{\rm eff}$
because they are optically-thick to their own reradiated FIR emission.  For surface densities
above $\sim5\times10^{3}$\,M$_\odot$ pc$^{-2}$, Galactic dust-to-gas ratio, and gas fractions
at birth of order unity, the characteristic temperature of star clusters exceeds the ice line
temperature (Figure \ref{figure:temp}; eqs.~\ref{tchot}, \ref{sigmaice}).   See the 
discussion of uncertainties and caveats in Section \ref{section:uncertain}. 

3.~Protoplanetary disks in such clusters are strongly isotropically irradiated,
and this cluster irradiation dominates the thermodynamics of both passive and active disks around Solar-type 
stars at semi-major axes beyond $\sim1-5$\,AU, changing the disk 
structure (Figure \ref{figure:cg}; 
eqs.~\ref{acritirrhot}-\ref{acritirreff} and \ref{acritaccthick} \& \ref{acritaccthin}).

4.~In the most massive dense star clusters, the temperature of protoplanetary disks 
should exceed the temperature for water ice condensation at $\sim150-170$\,K (Figures \ref{figure:t} \&
\ref{figure:tgc}) while the cluster is embedded.  This decreases the total 
amount of condensable material by a factor of $\sim2-5$, prohibiting formation
of gas- and ice-giant planets if they form by core accretion (Section \ref{section:core}),
and prohibiting habitability.  

5.~Planet formation by gravitational instability will also be suppressed in
protoplanetary disks within clusters because Toomre's $Q>1$.  
Typical disk masses that could become gravitationally 
unstable are required to exceed $\sim50-100$ times the Minimum Mass Solar Nebula
(Figure \ref{figure:tc}; Section \ref{section:grav}; eqs.~\ref{grav1}-\ref{grav3}).

6.~In these highly-irradiated environments, the overall 
accretion rate should be enhanced, leading to shorter disk lifetimes and less time to 
build planets.  The migration rates of planets in gas or planetesimal disks are also 
likely to be influenced (Section \ref{section:accretion}).

7.~Even accounting for their low metallicities, 
47 Tuc and several other Galactic globular clusters plausibly exceeded the
limiting surface density for their central regions to exceed $T_{\rm Ice}$,
and this may be a factor in explaining why 47 Tuc appears to have a paucity of
hot Jupiters (Figure \ref{figure:tgc}; Section \ref{section:globs}).  
A test of this scenario would be an RV/transit survey of NGC 6366;
with a metallicity similar to 47 Tuc, but a much 
lower surface density, it should have abundant giant planets.
Conversely, the relatively {\it more metal-rich}
globulars NGC 6440, 6441, or 6388, which had even higher temperatures than 
47 Tuc at birth should be devoid of planets (Figure \ref{figure:tgc}; Section \ref{section:rvcluster};
related discussion of NGC 6121).  A mutual comparison between planet
populations in clusters of different metallicities, but at fixed surface density,
and of different surface densities, but at fixed metallicity,  would
be very valuable. 

8.~Rapidly star-forming galaxies also maintain very high temperatures.
Those with average star formation rate surface densities 
$\gtrsim10^3$\,M$_\odot$ kpc$^{-2}$ yr$^{-1}$ have central temperatures 
that approach $T_{\rm Ice}$, and should thus globally suppress
planet formation (eq.~\ref{tgal}).  AGN activity might too.
Simple estimates suggest that the quiescent compact high-$z$ galaxies plausibly attained
temperatures exceeding $T_{\rm Ice}$ during formation (Figure \ref{figure:van}).  
The idea that these systems might be completely devoid of giant planets and habitable
planets is startling.
Because these systems may be the cores of today's elliptical galaxies, and because a 
component of the Galactic bulge and M31's bugle may have formed in such an 
extreme starburst, one generically expects future surveys of the planet populations
in these environments (e.g., via microlensing) to show a paucity of giant planets 
per star, even though (and, indeed, because) these systems are metal-rich.

9.~Finally, although the estimate is uncertain, between $\sim5-50$\% of the $z\sim0$ stellar
mass density formed in a star cluster with temperature exceeding $T_{\rm Ice}$, and thus
our expectations for the inventory of planets, and their incidence per star as a function of 
birth environment and metallicity may be significantly impacted.

\section*{Acknowledgments}
I gratefully acknowledge helpful conversations with Scott Gaudi at an early stage in this project.  
I thank Kris Stanek, Andy Gould, and Tony Piro for encouragement, and Kaitlin Kratter, 
Chris Kochanek, Scott Gaudi, and Fred Adams for a critical reading of the text.
This work is dedicated to my grandpa, Alan T.~Thompson.


\end{document}